\documentclass[a4paper,fleqn]{cas-dc}

\usepackage[numbers,sort&compress]{natbib}
\usepackage{amsmath,amsfonts,amssymb}
\usepackage{graphicx}
\usepackage{array}
\usepackage{booktabs}
\usepackage{url}
\usepackage{subcaption}
\usepackage{textcomp}
\usepackage{float}
\usepackage[normalem]{ulem}
\usepackage[american]{circuitikz}
\usetikzlibrary{calc,positioning,arrows.meta}
\ctikzset{bipoles/length=1.05cm}

\makeatletter
\@ifundefined{pgfk@/tikz/block lateral anchors pos}{%
  \pgfkeys{/tikz/block lateral anchors pos/.code={}}%
}{}
\@ifundefined{pgfk@/tikz/block right anchors pos}{%
  \pgfkeys{/tikz/block right anchors pos/.code={}}%
}{}
\@ifundefined{pgfk@/tikz/circuitikz/block right anchors pos}{%
  \pgfkeys{/tikz/circuitikz/block right anchors pos/.code={}}%
}{}
\makeatother

\def\tsc#1{\csdef{#1}{\textsc{\lowercase{#1}}\xspace}}
\tsc{PFSCR}
\tsc{CLLC}
\tsc{GVM}
\tsc{CPO}
\tsc{ORBITRON}
\tsc{MCX}
\tsc{CMFX}
\tsc{PSP}

\begin{document}
\let\WriteBookmarks\relax
\restylefloat{figure}

\shorttitle{Conceptual Design of a 5 MV High-Voltage System for a Compact Orbitron}
\shortauthors{M. Borghei et~al.}

\title[mode=title]{Conceptual Design of a 5 MV High-Voltage System for a Compact Orbitron Fusion Device}

\author[1]{Moein Borghei}[orcid=0000-0002-4188-432X]
\cormark[1]
\ead{mborghei@avalanche.energy}
\credit{Conceptualization, Investigation, Methodology, Project administration, Visualization, Writing -- Original Draft, Writing -- Review \& Editing}

\affiliation[1]{organization={Avalanche Energy Designs},
            addressline={9100 E Marginal Way S},
            city={Tukwila},
            postcode={98108},
            state={WA},
            country={USA}}

\author[1]{Robin Langtry}
\credit{Investigation, Methodology, Writing -- Review \& Editing}

\author[1]{Nicola Van Moon}
\credit{Methodology, Writing -- Review \& Editing}

\author[1]{Matthew Affolter}
\credit{Writing -- Review \& Editing}

\author[1]{Erik McKee}
\credit{Writing -- Review \& Editing}

\cortext[1]{Corresponding author}

\begin{abstract}
The high-voltage (HV) system is a key part of the puzzle toward achieving net-positive fusion power in a compact Orbitron device. This paper presents a conceptual engineering design for a 5 MV, 20 mA, 100 kW HV system for a next-generation Orbitron, serving as an upper-bound design point for the HV system. The MC-TRANS++ 0-D centrifugal-mirror code is used to set the load requirements: a 1 m long, 0.45 m first-wall-radius central cell, 2.5 T on-axis and 10 T mirror-throat fields, and a 5 MV radial potential drop sustaining supersonic rotation ($M\approx 4.8$). This delivers a triple product of $5.8\times10^{21}$~keV\,s\,m$^{-3}$ and $Q_{\text{sci}}\approx 8.5$. The HV power supply integrates an AC–DC converter, a SiC full-bridge inverter, a CLLC resonant network with an air-core step-up transformer, and feeds a 120-stage parallel-fed series-cascaded rectifier (PFSCR) operating at 250 kHz. At full load, this architecture yields an end-to-end efficiency of 85\%, substantially higher than the $\sim$60\% typical of triode-driven MV-class power supplies. The 5 MV voltage is delivered to a radially nested 50-gap bushing system that interfaces pressurized perfluoroketone/CO2/O2 insulating gas to high vacuum, using brazed alumina spacers with a corrugated profile. The design draws directly on the measured behavior of a previously demonstrated 300 kV prototype (the Hammerhead bushing), from which six quantitative lessons are extracted and mapped onto the 5 MV design. Electrostatic simulations of the complete assembly at rated voltage show maximum surface fields of 30.3 MV/m on the corona rings, 7.7 MV/m at the cathode triple junction, and average vacuum-gap fields ranging from 5.5 to 40 MV/m. These field strengths remain below breakdown thresholds due to the implementation of triple-junction shielding, insulation corrugation, and axial magnetic insulation at 2.5-10 T. The conceptual system has a form factor of approximately 2.7 meters in diameter and 6.5 meters in length. Based on projected plasma performance, the system delivers a net power output of 86--171 kW\textsubscript{e}.
\end{abstract}


\maketitle

\section{Introduction}\label{sec:intro}
Global electricity demand has grown faster over the past two years than in any comparable window since the early 2010s, driven by electrification of transport, data-center expansion, and renewed industrial activity. The International Energy Agency reports a 2.2\% increase in demand in 2024, nearly double the 2013--2023 average of $\sim$1.3\%~\cite{IEA_Electricity2025}, and McKinsey projects that global electricity demand will more than double by 2050~\cite{McKinsey_GlobalEnergyPerspective2025}. In response, private and public investment in fusion energy has accelerated, with over 50 companies now commanding nearly \$10\,B in disclosed funding~\cite{FIA_GlobalFusionIndustryReport_2025}.

High-voltage (HV) subsystems are essential to several fusion concepts. Prominent examples include HV accelerators for neutral-beam injection~\cite{watanabe_development_1998}, HV cathodes for inertial-electrostatic confinement~\cite{hirsch_inertial-electrostatic_1967,fancher_fusion_2018}, HV power conditioning for inertial-confinement-fusion lasers~\cite{arnold_power_2015}, and pulsed-power drivers for Z-pinch fusion~\cite{sinars_review_2020}. The Orbitron, a hybrid magneto-electric confinement device~\cite{affolter_orbitron_2024}, relies on a central cathode held at a high negative DC potential to produce a deep electrostatic well that accelerates ions to fusion-relevant energies, emits primary electrons, and assists with plasma formation. In the classical Orbitron, electrons are magnetized by an axial field while the ions are largely unmagnetized.

Raising the axial field drives the Orbitron toward a centrifugal-mirror-like regime: the ions also become subject to $\mathbf{E}\times\mathbf{B}$ rotation, their orbits shrink, and the resulting supersonic rotation reduces axial mirror losses and shear-stabilizes low-frequency interchange modes~\cite{ellis_steady_2005,bekhtenev_problems_1980}. The trade-off is that smaller orbits sample smaller portions of the cathode potential, which reduces electrostatic heating efficiency. Compensating for this loss and supplying the rotational power against viscous and parallel losses sets the terminal-voltage requirement of the HV supply firmly in the multi-megavolt range for a net-energy device. The feasibility of centrifugal confinement has been explored across several decades: the Plasma in Crossed Fields (PSP)\footnote{PSP is an English transliteration of the Russian acronym, which translates literally to Plasma in Skreshchennykh (Crossed) Polyakh (Fields).} experiment at the Budker Institute used segmented ring electrodes in the mirror region with voltages up to $\sim$450~kV, producing stable rotating plasmas with ion energies of $\sim$20~keV at densities near $10^{18}$~m\textsuperscript{-3}~\cite{abdrashitov_hot_1991,volosov_mhd_2009}; the Maryland Centrifugal Experiment (MCX) demonstrated routine supersonic rotation at Mach 2--2.5 with fully-ionized plasmas at $n_e \gtrsim 10^{20}$~m\textsuperscript{-3}~\cite{ellis_steady_2005,lunsford_parametric_2007}; and the Centrifugal Mirror Fusion Experiment (CMFX) at the University of Maryland now uses superconducting magnets and insulating end-plates, with a DC cathode bias up to $\sim$100~kV (currently limited to $\sim$60~kV by end-insulator discharges) to drive $\mathbf{E}\times\mathbf{B}$ rotation in a deuterium plasma~\cite{ball_measurements_2025,schwartz_mctrans_2024}. These experiments establish the physics case; the remaining barrier between them and a net-positive device is an engineering one --- the generation, regulation, and vacuum delivery of a DC potential more than an order of magnitude higher.

Existing MV-class DC sources illustrate the nature of that gap. Electrostatic generators such as Van de Graaff and Pelletron machines can reach 25~MV, but their reliance on mechanical charge transport limits the delivered current to well below 1~mA~\cite{flugge_van_1959,hinterberger_electrostatic_2006,herb_pelletron_1971}. Conventional single-source Cockcroft--Walton (Greinacher) rectifiers suffer from voltage drop and ripple that scale with the cube and square of the stage count respectively~\cite{kuffel_high_2000}, which makes tens of milliamperes at multi-megavolt scales impractical. Insulating-core transformers (ICTs)~\cite{van_de_graaff_high_1965,frost_investigation_2019}, cascaded transformers~\cite{kuffel_high_2000}, and parallel-fed capacitively-coupled rectifiers (the Dynamitron class~\cite{cleland_dynamitrons_1965}) overcome these limits by distributing the multiplication across multiple AC sources or magnetic couplings. Of these, the Dynamitron has been proven commercially at current levels compatible with the Orbitron load, but its classical implementation uses vacuum triodes in Class C operation~\cite{thompson_design_1969,thompson_high-power_1989,galloway_new_2004,matsuyama_improvement_2013}, which caps system efficiency at roughly 60\%~\cite{galloway_new_2004-1}, imposes 120 Hz terminal ripple~\cite{thompson_multi-loop_1967}, and limits tube lifetime to 20,000--25,000 hours~\cite{galloway_new_2004}. Solid-state oscillator replacements have been proposed, and a 1.3~MV implementation has been reported~\cite{gong_solid-state_2024}, but no demonstrated 5~MV-class solid-state Dynamitron exists in the hundreds-of-kilowatt range.

\subsection{Scope and contributions}
This paper presents the conceptual engineering design of an integrated 5 MV HV system for a next-generation Orbitron. This work draws on lessons from a previously demonstrated 300 kV vacuum bushing prototype \cite{borghei_hammerhead_2025} and lays the foundation for the experimental implementation. The main contributions are summarized as follows:
\begin{enumerate}[(1)]
    \item The design and parameter identification of a 100 kW HV generator that pairs a SiC-MOSFET CLLC resonant tank with a 120-stage parallel-fed series-cascaded rectifier, reaching $>$93\% DC--DC efficiency and $\sim$85\% end-to-end efficiency (Section~\ref{sec:hv_gen}).
    \item A compact vacuum bushing topology that transmits 5 MV from a pressurized C\textsubscript{5}F\textsubscript{10}O / CO\textsubscript{2} / O\textsubscript{2} gas medium to high vacuum through 50 brazed alumina spacers, with triple-junction shielding inherited from the Hammerhead prototype (Section~\ref{sec:hv_deliv}).
    \item An end-electrode system capable of imposing a tailored radial potential profile on the rotating plasma, electrically and mechanically integrated into the bushing (Section~\ref{sec:hv_deliv}).
\end{enumerate}

The remainder of the paper is organized as follows. Section~\ref{sec:system} summarizes the Orbitron, its load requirements, and includes a discussion of magnetic insulation and stability. Section~\ref{sec:lessons} summarizes the quantitative outcomes of the 300 kV prototype campaign and their implications for the 5 MV design. Section~\ref{sec:hv_gen} details the HV generation chain, including a new analysis of high-frequency dielectric effects and a commercial-availability assessment of passive components. Section~\ref{sec:hv_deliv} describes the HV delivery architecture -- the gas vessel, vacuum bushing, and Orbitron main cell. Section~\ref{sec:conclusion} concludes and outlines the scope of future work.

\section{The Orbitron, Requirements and Plasma Load}\label{sec:system}

\subsection{Conceptual view of the Orbitron machine}\label{sec:machine_overview}
\begin{figure}[h]
    \centering
    \includegraphics[width=0.5\textwidth]{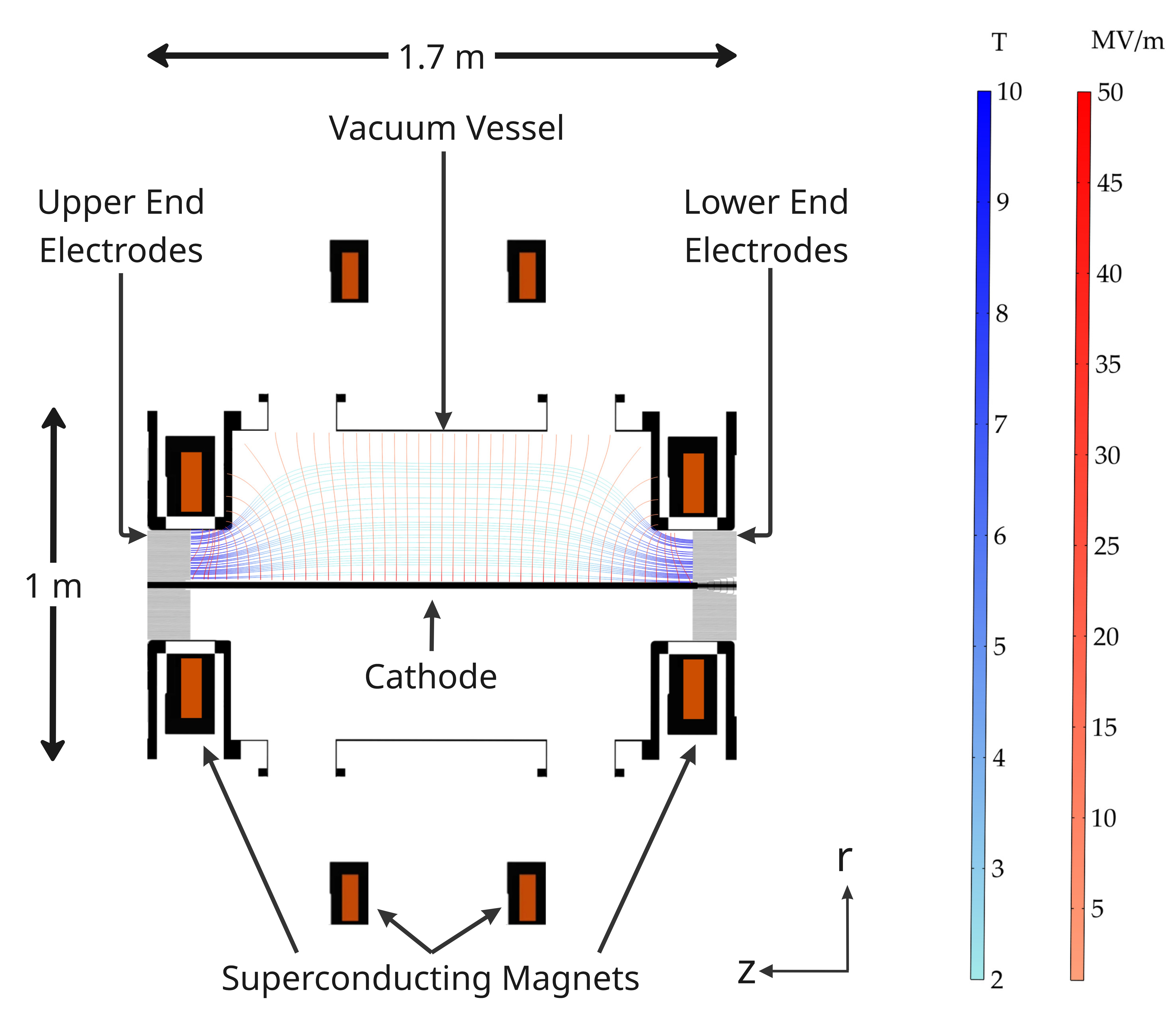}
    \caption{Conceptual schematic of the presented Orbitron, with the magnetic flux density (blue) and electric field distribution (red) highlighted.}
    \label{fig:orbitron_concept}
\end{figure}

Figure~\ref{fig:orbitron_concept} provides a conceptual view of the Orbitron device showing the relative placement of the central cathode, grounded anode (vacuum-vessel wall), end electrodes, magnetic-field lines, and the direction of $\mathbf{E}\times\mathbf{B}$ rotation. The HV system described in the following sections interfaces with the machine at the left-hand end, where the bushing brings the 5 MV DC potential from the pressurized gas vessel into the vacuum chamber and feeds the central cathode and 49 additional voltage-graded electrodes.

Figure~\ref{fig:orbitron_concept} illustrates the crossed magnetic and electric field lines within the device. The electric field is established by a coaxial electrode configuration, with the central cathode and the vacuum vessel acting as the anode. This field can be further tailored using two sets of end electrodes positioned within the end magnet bores. Regarding the magnetic field, steady-state operation at 2.5~T on-axis and 10~T at the mirror throats is a primary engineering driver; consequently, the design utilizes a high-temperature superconducting (HTS) coil set cooled by forced-flow helium. While the detailed thermal design of these coils is beyond the scope of this paper, their conceptual placement is shown in Figure~\ref{fig:orbitron_concept}. The interaction of these crossed fields induces rapid plasma rotation around the central electrode, generating the centrifugal force required to effectively confine the plasma, particularly during supersonic operation.

\subsection{Load specification from 0-D plasma modeling}\label{sec:5MV_required}
The plasma load considered in this study is determined from a steady-state deuterium-tritium (D-T) solution obtained using MC-TRANS++, a 0-D centrifugal-mirror plasma model developed at the University of Maryland \cite{schwartz_mctrans_2024}. For a prescribed device geometry, magnetic-field configuration, fuel composition, and operating conditions, the model solves the particle- and energy-balance equations to estimate the steady-state plasma properties and fusion performance. The temperatures, confinement times, fusion triple product, fusion power, and scientific energy gain reported below are therefore model predictions rather than experimentally validated performance of the proposed 5 MV Orbitron. Tables \ref{tab:mctrans_device} and \ref{tab:mctrans_performance} summarize the prescribed device parameters and the corresponding calculated plasma-performance quantities.

 The modeled cell has a length of 1~m and a first-wall radius of 0.45~m. Magnetic confinement and gross stability require 2.5~T on axis at the device center and 10~T at the mirror throat. Under these conditions the steady-state solution yields ion and electron temperatures of 45~keV and 55~keV, each species at densities near $4\times10^{19}$~m\textsuperscript{-3}, a fusion triple product of $nT\tau \approx 5.8\times10^{21}$~keV\,s\,m\textsuperscript{-3}, and a scientific gain $Q_{\text{sci}}\approx 8.5$ with 0.434~MW of thermal output from a plasma volume of 0.282~m\textsuperscript{3}.

Centrifugal confinement requires supersonic rotation to reduce axial losses; the solution predicts Mach 4.76, corresponding to an azimuthal velocity of $6.9\times 10^{6}$~m/s. Sustaining this rotation against viscous and parallel losses ($\sim$35~kW) requires a strong radial electric field and hence a 5 MV radial potential drop in steady state. Electrically, the plasma acts as a high-impedance load of 623~M$\Omega$ in parallel with 89.9~pF. Allowing margin for leakage, the HV supply is specified at 20~mA at 5~MV (100~kW delivered to the plasma). These load characteristics set the primary requirements for the HV system described in the subsequent sections.

\begin{table}[t]
\caption{Compact Orbitron device and plasma parameters from the
         MC-TRANS++ steady-state D--T solution.}
\label{tab:mctrans_device}
\centering
\begin{tabular}{@{}llc@{}}
\toprule
Parameter & Symbol & Value \\
\midrule
\multicolumn{3}{@{}l}{\emph{Geometry and configuration}}\\
Fuel species              & ---            & D--T \\
Central-cell length       & $L$            & 1.0 m \\
First-wall radius         & $R_\text{wall}$ & 0.45 m \\
Plasma inner radius       & $R_\text{min}$ & 0.01 m \\
Plasma outer radius       & $R_\text{max}$ & 0.30 m \\
Outer radius at throat    & ---            & 0.15 m \\
Exhaust radius            & $R_\text{exh}$ & 0.04 m \\
Plasma volume             & $V$            & 0.282 m\textsuperscript{3} \\[2pt]
\multicolumn{3}{@{}l}{\emph{Magnetic configuration}}\\
Central-cell field        & $B_0$          & 2.5 T \\
Mirror-throat field       & $B_\text{m}$   & 10 T \\
Mirror ratio              & $R_\text{m}$   & 4 \\[2pt]
\multicolumn{3}{@{}l}{\emph{Plasma parameters}}\\
Electron density          & $n_e$          & $4\times10^{19}$ m\textsuperscript{-3} \\
Ion density               & $n_i$          & $4\times10^{19}$ m\textsuperscript{-3} \\
Electron temperature      & $T_e$          & 54.8 keV \\
Ion temperature           & $T_i$          & 45.3 keV \\
Effective charge          & $Z_\text{eff}$ & 1.0 \\
Plasma beta               & $\beta$        & 25.8\% \\
Ion Larmor radius         & $\rho_i$       & 0.0137 m \\
Normalized gyroradius     & $\rho^*$       & 0.0947 \\
Ion collisionality        & $\nu^*$        & $2.24\times10^{-6}$ \\
Operating Mach number     & $M$            & 4.76 \\
Alfv\'en Mach number      & $M_A$          & 1.26 \\
Rotation velocity         & $v$            & $6.9\times10^{6}$ m/s \\
Angular velocity          & $\omega$       & $4.45\times10^{7}$ s\textsuperscript{-1} \\
\bottomrule
\end{tabular}
\end{table}

\begin{table}[t]
\caption{Compact Orbitron performance metrics and figures of merit
         from the MC-TRANS++ steady-state D--T solution.}
\label{tab:mctrans_performance}
\centering
\begin{tabular}{@{}llc@{}}
\toprule
Metric & Symbol & Value \\
\midrule
\multicolumn{3}{@{}l}{\emph{Electrical load and power balance}}\\
Total potential drop          & $V_\text{0}$ & 5 MV \\
Radial current drawn          & $I$           & 8.03 mA \\
Power to sustain rotation     & $P_\text{rot}$ & 40.1 kW \\
Viscous heating               & ---           & 25.3 kW \\
Total heat losses             & $P_\text{loss}$ & 84.2 kW \\
Plasma resistance             & $R_p$         & 623 M$\Omega$ \\
Plasma capacitance            & $C_p$         & 89.9 pF \\[2pt]
\multicolumn{3}{@{}l}{\emph{Energy confinement}}\\
Total stored energy           & $W$           & 1.39 MJ \\
\quad rotational kinetic      & ---           & 1.12 MJ \\
\quad thermal                 & ---           & 0.271 MJ \\
Energy confinement time       & $\tau_E$      & 3.22 s \\
Particle confinement time     & $\tau_p$      & 14.5 s \\
Ion--electron equilibration   & $\tau_\text{eq}$ & 8.76 s \\[2pt]
\multicolumn{3}{@{}l}{\emph{Fusion output and figures of merit}}\\
Fusion triple product         & $n T \tau_E$  & $5.83\times10^{21}$ keV\,s\,m\textsuperscript{-3} \\
Alpha heating power           & $P_\alpha$    & 59.1 kW \\
Fusion power (alphas)         & ---           & 68.2 kW \\
Fusion power (neutrons)       & ---           & 273 kW \\
Neutron Wall Loading & --- & 66.4 kW/m\textsuperscript{2} \\
Total thermal power           & $P_\text{th}$ & 433 kW \\
Scientific gain               & $Q_\text{sci}$ & 8.5 \\
\bottomrule
\end{tabular}
\end{table}

\subsection{Magnetic Insulation of Vacuum Gap}\label{sec:magnetization}
Because the Orbitron operates at very high voltages, the required clearance for vacuum insulation between the central conductor and the grounded chamber at the device center would be several meters to allow electrode conditioning. To avoid this and keep the Orbitron compact, the axial magnetic field must be used to provide magnetic insulation and prevent vacuum breakdown. The applied magnetic field must equal or exceed the critical magnetic field (often called the Hull cutoff field) to ensure that electrons emitted from the cathode are deflected back toward it rather than crossing the gap to the anode. Transverse electron transport across field lines is therefore strongly inhibited, and an $\mathbf{E}\times\mathbf{B}$ insulation regime applies~\cite{lovelace_theory_1974,liziakin_generation_2024}.

For the cases where electron velocities approach the speed of light (as is the case in this study), the critical axial magnetic field $B_c$ required to insulate concentric cylinders is given by the relativistic Hull cutoff equation \cite{creedon_magnetic_1977}:

\begin{equation}
B_c = \frac{2R_2 m_e c}{e(R_2^2 - R_1^2)} \left[ \left(\frac{eV_0}{m_e c^2}\right)^2 + 2\frac{eV_0}{m_e c^2} \right]^{1/2}
\end{equation}
Where $R_2$ is the radius of the outer electrode (anode), $R_1$ is the radius of the inner electrode (cathode), $V_0$ is the applied electric potential, $ c$ is the speed of light, and $m_e$ and $e$ are the electron rest mass and charge, respectively.  

Note that the central electrode radius is additionally constrained by plasma-facing material lifetime. Bombardment by energetic plasma particles, together with localized field-emission or discharge events, can produce sputtering and surface recession of the electrode. Therefore, the \(R_1\) value should be treated as a conceptual design input rather than a lifetime-qualified dimension. Determination of the minimum practical cathode radius requires a coupled estimate of incident particle flux and energy, sputtering yield, thermal loading, and allowable recession over the target operating lifetime. The assessment is reserved for the future design and experimental qualification phase.

To provide a sufficient magnetic-insulation margin, practical designs often apply a magnetic field that is significantly higher than the theoretical limit, typically operating at least 1.5 to 3.0 times above the critical Hull cutoff field.

For the proposed device's geometry and operating voltage, $B_c=0.25 \ T$. Therefore, the axial magnetic field must be much greater. This condition is satisfied, as the minimum magnetic field inside the vacuum chamber is 2.5 T at the center.

\subsection{MHD and high-velocity instability considerations}\label{sec:mhd}
Rotating mirror plasmas are known to be prone to two classes of low-frequency instabilities: Kelvin--Helmholtz-like modes driven by the radial shear of the azimuthal velocity, and interchange modes driven by the centrifugal effective gravity in the curvature of the magnetic field~\cite{bekhtenev_problems_1980,ellis_steady_2005,huang_2001}. The $\mathbf{E}\times\mathbf{B}$ configuration imposed by the cathode and the end-plugging electrodes mitigates these in two ways. First, the sheared $\mathbf{E}\times\mathbf{B}$ flow itself provides velocity-shear stabilization of interchange modes once the Mach number exceeds unity, as demonstrated experimentally in PSP-2~\cite{abdrashitov_hot_1991,volosov_mhd_2009} and MCX~\cite{ellis_steady_2005,lunsford_parametric_2007}. Second, the end-plugging electrodes (Section~\ref{sec:hv_deliv}) allow the imposed radial potential profile $\phi(r)$ to be tailored, which in turn sets the local shear $d\omega/dr$ where $\omega=-(1/rB)\,d\phi/dr$ is the rotation frequency. Tuning this profile both shapes the Mach number distribution and suppresses the resonant Kelvin--Helmholtz band. MC-TRANS++ solutions reported in Section~\ref{sec:5MV_required} assume a profile that is MHD-stable under these criteria; a detailed transient MHD analysis is outside the scope of this engineering study and will be reported separately.

\section{Inputs from the 300 kV Hammerhead Bushing}\label{sec:lessons}
The present 5 MV design draws on multiple inputs from the design, simulation, and experimental work on the 300 kV compact vacuum bushing, \emph{Hammerhead}, which the authors recently reported and characterized experimentally over 144~h of stable operation at 300~kV with dark current below 10~$\mu$A~\cite{borghei_hammerhead_2025}. Table~\ref{tab:hh_lessons} summarizes six quantitative outcomes of that prototype campaign and the specific 5 MV design choice each one informs.

\begin{table*}[t]
\caption{Lessons from the 300 kV Hammerhead prototype and their application to the 5 MV design.}
\label{tab:hh_lessons}
\centering
\renewcommand{\arraystretch}{1.15}
\begin{tabular}{@{}p{4.2cm}p{5.2cm}p{6.5cm}@{}}
\toprule
\textbf{Observation at 300 kV} & \textbf{Quantitative outcome} & \textbf{Implication for 5 MV design} \\
\midrule
Equipotential-well shielding of the cathode triple junction & Simulated triple-junction field reduced from tens of MV/m to 1.04~MV/m; no flashover initiation in 144~h~\cite{borghei_hammerhead_2025} & Each of the 50 electrode gaps in the 5 MV bushing uses the same shielded triple-junction geometry; simulated field at the 5 MV cathode triple junction is 7.7~MV/m, still below the 40~MV/m design cap for switch-on on tungsten~\cite{furuta_reduction_2005} \\
Corrugated (sawtooth) MACOR insulator geometry & Flashover path extended to $>$400~mm along a 10~cm axial insulator; electrons on the surface forced to climb against the applied field & Each 100 kV stage uses a corrugated Al\textsubscript{2}O\textsubscript{3} insulator with a 210 mm flashover path. Countersink feature of the first fin is retained \\
Refractory-metal cathode (molybdenum) & Surface field held below 40 MV/m, a factor of 0.6 below the reported switch-on field~\cite{furuta_reduction_2005} & 5 MV cathode and electrodes switch to tungsten for brazeability with alumina and equivalent work function; peak field on the 5 MV rings capped at 30.3 MV/m \\
Permittivity matching across solid / gas / vacuum interfaces & Materials chosen with $\varepsilon_r\in[2.8,6]$ to limit field discontinuities at the triple junction & 5 MV interfaces match 97\% Al\textsubscript{2}O\textsubscript{3} ($\varepsilon_r\approx 9.4$) to the C\textsubscript{5}F\textsubscript{10}O mixture on the gas side and vacuum ($\varepsilon_r=1$) on the plasma side, with corona-ring geometry reducing the local field enhancement factor to $<2$ \\
Particulate (Cu/Ag) contamination as the main de-conditioning cause & Copper from seals and silver from fastener plating observed in SEM/EDS as re-deposited clumps, requiring multi-day recovery & In the 5 MV assembly, all copper seals are removed from the vacuum envelope; fasteners near active electrodes use uncoated stainless; cleanroom class is upgraded from class 10,000 to class 1,000 \\
Stored-energy budget for safe conditioning & $\sim$195 J at 300 kV in the full test stand was sufficient for upward conditioning without electrode damage & The 5 MV PFSCR ladder stores $<$100 J at full voltage~\cite{galloway_new_2004-1}. The RF choke $L_i$ and the corona-ring capacitance network are sized so that no single fault path discharges more than $\sim$30 J, well below the Hammerhead-validated upper bound \\
\bottomrule
\end{tabular}
\end{table*}

These lessons serve as the foundation for the subsequent subsystem descriptions. While the 300 kV experiment relied on a single gap to withstand the entire potential, our 5 MV design distributes the load across 50 discrete 100 kV gaps. Consequently, the design requirements established by the Hammerhead prototype act as conservative benchmarks for each individual voltage stage in the 5 MV system.

\begin{figure*}[b]
\centering
\resizebox{\textwidth}{!}{%
\begin{tikzpicture}[
  font=\small,
  box/.style={draw, rounded corners, align=center, minimum width=1cm, minimum height=1.0cm},
  arrow/.style={-{Latex[length=2mm]}, thick},
  node distance=2cm
]
\node[draw, circle, minimum size=1.1cm, align=center] (src) {$3\phi$\\AC};
\node[box, right= of src] (b1) {\textbf{AC$\rightarrow$DC}};
\node[box, right=of b1] (b2) {\textbf{DC$\rightarrow$AC}};
\node[box, right=of b2] (b3) {\textbf{Resonant Tank}};
\node[box, right=of b3] (b4) {\textbf{PFSCR}};
\node[box, below=1.2cm of b2] (b5) {\textbf{Feedback \& control}};
\node[box, right=of b4] (orb1) {\textbf{Vacuum Bushing}};
\node[draw, circle, minimum size=1.1cm, align=center, right=of orb1] (orb2) {\textbf{Orbitron}};
\draw (16.5,0) node[above] {5 MV};
\draw (16.5,-0.5) node[above] {DC};
\draw (13,0) node[above] {$150\,\mathrm{kV}_{\mathrm{pk}}$};
\draw (13,-0.5) node[above] {250 kHz};
\draw (8.5,0) node[above] {$0.5$--$1\,\mathrm{kV}_{\mathrm{pk}}$};
\draw (8.5,-0.5) node[above] {250 kHz};
\draw (5,0) node[above] {1 kV};
\draw (5,-0.5) node[above] {DC};
\draw (1.5,0) node[above] {$480\,\mathrm{V}_{\mathrm{rms}}$};
\draw (1.5,-0.5) node[above] {60 Hz};
\draw[arrow] (src) -- (b1);
\draw[arrow] (b1) -- (b2);
\draw[arrow] (b2) -- (b3);
\draw[arrow] (b3) -- (b4);
\draw[arrow] (b4) -- (orb1);
\draw[arrow] (orb1) -- (orb2);
\draw[arrow] (15.75,0) |- (b5.east);
\draw[arrow] (b5.west) -| (b1.south);
\end{tikzpicture}
}
\caption{High-level block diagram of the 5 MV DC generation and delivery chain with closed-loop regulation.}
\label{fig:hv_chain_block_diagram}
\end{figure*}
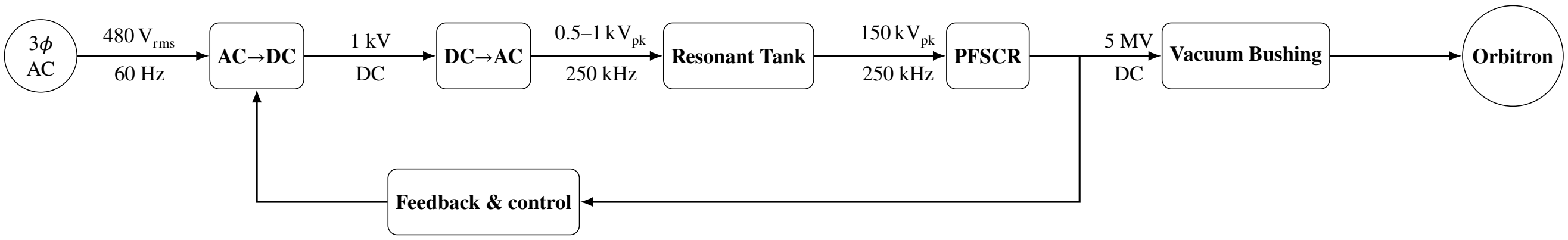

\section{HV Generation}\label{sec:hv_gen}
\subsection{Overall architecture}
The high-level architecture for generating and delivering up to 5 MV to the Orbitron is shown in Figure~\ref{fig:hv_chain_block_diagram}. The chain begins with a 480~V\textsubscript{rms}, 60~Hz, 3-phase grid input; a commercial AC--DC front end produces a regulated 1 kV DC bus; a SiC full-bridge inverter converts this to a 0.5--1 kV\textsubscript{pk} square wave at 250 kHz; a CLLC resonant tank with an air-core step-up transformer provides a 150 kV\textsubscript{pk} sinusoidal drive; and a 120-stage parallel-fed series-cascaded rectifier (PFSCR) multiplies that up to the 5 MV DC terminal. A regulation loop compares the measured terminal voltage against the setpoint and closes through the AC--DC stage. Sections~\ref{sec:pfscr}--\ref{sec:sensing} describe each block.

\subsection{Parallel-fed series-cascaded rectifier}\label{sec:pfscr}
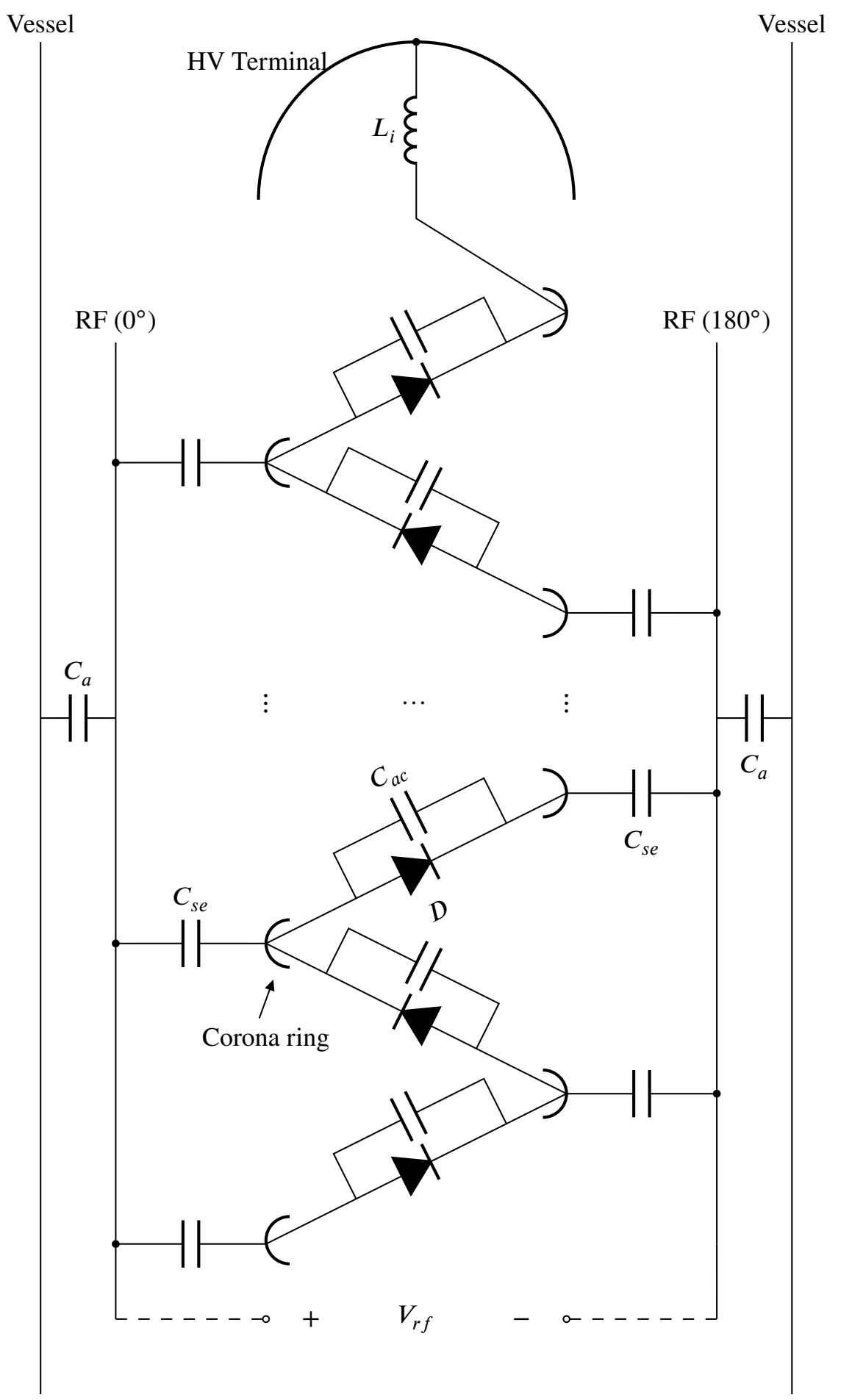
\begin{figure}[h]
\centering
\resizebox{3in}{!}{%
    \begin{circuitikz}[scale=0.7, transform shape, >=latex]
    \def\yStep{4}
    \def\yHalf{2}
    \def\xWallLeft{0}
    \def\xRFLeft{1}
    \def\xNodeLeft{3}
    \def\xNodeRight{7}
    \def\xRFRight{9}
    \def\xWallRight{10}
    \draw (\xWallLeft, -2) -- (\xWallLeft, {3.5*\yStep + \yHalf}) node[above] {Vessel};
    \draw (\xWallRight, -2) -- (\xWallRight, {3.5*\yStep + \yHalf}) node[above] {Vessel};
    \draw (\xWallLeft, 7) to[C,l={$C_a$}] (\xRFLeft, 7);
    \draw (\xWallRight, 7) to[C,l={$C_a$}] (\xRFRight, 7);
    \draw (\xRFLeft, 0) -- (\xRFLeft, {2.5*\yStep + \yHalf}) node[above] {RF ($0^\circ$)};
    \draw (\xRFRight, 0) -- (\xRFRight, {2.5*\yStep + \yHalf}) node[above] {RF ($180^\circ$)};
    \draw (\xRFLeft+2, -1) to[open, v=$V_{rf}$,o-o] (\xRFRight-2, -1);
    \draw (\xRFLeft, 0)-- (\xRFLeft, -1);
    \draw  (\xRFLeft, -1)  [dashed] -- (\xRFLeft+2, -1);
    \draw (\xRFRight, 0) to[short] (\xRFRight, -1);
    \draw  (\xRFRight, -1)  [dashed]-- (\xRFRight-2, -1);
    \draw (\xRFLeft, 0) to[C, *-] (\xNodeLeft, 0);
    \draw (\xRFRight, \yHalf) to[C, *-] (\xNodeRight, \yHalf);
    \draw (\xNodeLeft, 0) to[D*, -] (\xNodeRight, \yHalf);
    \draw (1.2+\xNodeLeft, 0.6) to[short] (0.9+\xNodeLeft, 1.2)
     to [C]  (2.9+\xNodeLeft, 2.2)
     to [short] (3.2+\xNodeLeft, 1.6);
    \draw (3,10.4) node[iecsocketL, scale=3] {};
    \draw (3,4) node[iecsocketL, scale=3] {};
    \draw (3,0.05) node[iecsocketL, scale=3] {};
    \draw (7,12.4) node[iecsocketL, scale=3, rotate=180] {};
    \draw (7,8.4) node[iecsocketL, scale=3, rotate=180] {};
    \draw (7,6) node[iecsocketL, scale=3, rotate=180] {};
    \draw (7,2) node[iecsocketL, scale=3, rotate=180] {};
    \draw (3,3.25) node[flowarrow, scale=1, rotate=70] {}
    node[below] at (3,3) {Corona ring};
    \draw (5,16) node[iecsocketL, scale=20, rotate=-90] {}
    node[left] at (3.95,15.75) {HV Terminal};
    \draw (\xRFLeft, \yStep) to[C, l=$C_{se}$, *-] (\xNodeLeft, \yStep);
    \draw (\xNodeRight, \yHalf) to[D*, -] (\xNodeLeft, \yStep);
    \draw (-1.2+\xNodeRight, \yHalf+0.6) to[short] (-0.9+\xNodeRight, \yHalf+1.2)
     to [C]  (-2.9+\xNodeRight, \yHalf+2.2)
     to [short] (-3.2+\xNodeRight, \yHalf+1.6);
    \draw (\xRFRight, {\yStep + \yHalf}) to[C, l=$C_{se}$, *-] (\xNodeRight, {\yStep + \yHalf});
    \draw (\xNodeLeft, \yStep) to[D*, l_=$D$, -] (\xNodeRight, {\yStep + \yHalf});
    \draw (1.2+\xNodeLeft, \yStep+0.6) to[short] (0.9+\xNodeLeft, \yStep+1.2)
     to [C,l=$C_{ac}$]  (2.9+\xNodeLeft, \yStep+2.2)
     to [short] (3.2+\xNodeLeft, \yStep+1.6);
    \node at (\xNodeLeft, {1.8*\yStep}) {$\vdots$};
    \node at ({(\xNodeLeft+\xNodeRight)/2}, {1.8*\yStep}) {$\dots$};
    \node at (\xNodeRight, {1.8*\yStep}) {$\vdots$};
    \def\yN{1.6*\yStep}
    \draw (\xRFLeft, {\yN + \yStep}) to[C, *-] (\xNodeLeft, {\yN + \yStep});
    \draw (\xRFRight, {\yN + \yHalf}) to[C, *-] (\xNodeRight, {\yN + \yHalf});
    \draw (\xNodeRight, \yN + \yHalf) to[D*, -] (\xNodeLeft, {\yN + \yStep});
    \draw (-1.2+\xNodeRight, {\yN + \yHalf+0.6}) to[short] (-0.9+\xNodeRight, {\yN + \yHalf+1.2})
     to [C]  (-2.9+\xNodeRight, {\yN + \yHalf+2.2})
     to [short] (-3.2+\xNodeRight, {\yN + \yHalf+1.6});
    \draw (\xNodeLeft, {\yN + \yStep}) to[D*,-] (\xNodeRight, {\yN +\yStep + \yHalf});
    \draw (1.2+\xNodeLeft, \yN +\yStep +0.6) to[short] (0.9+\xNodeLeft, \yN +\yStep+1.2)
     to [C]  (2.9+\xNodeLeft, \yN +\yStep+2.2)
     to [short] (3.2+\xNodeLeft, \yN +\yStep+1.6);
    \draw (\xNodeRight+0.01, {\yN +\yStep + \yHalf}) to[short] (\xNodeRight-2, {\yN +\yStep + \yHalf+1.25}) to[L,-*, l=$L_i$] ++(0,2.35); 
    \end{circuitikz}
}
\caption{Symmetrical parallel-fed series-cascaded rectifier column.}
\label{fig:pfscr_topology}
\end{figure}

The parallel-fed series-cascaded rectifier (PFSCR) topology adopted here is best known through its commercial Dynamitron implementation~\cite{cleland_dynamitrons_1965}. An RF oscillator feeds two semi-cylindrical drive electrodes (the RF electrodes) that enclose the high-voltage rectifier assembly~\cite{cleland_new_1993,galloway_new_2004-1}. Semi-circular corona rings attached to each rectifier stage (historically vacuum diodes, now solid-state stacks) form a ladder of electrodes along the acceleration column. The oscillating potential between the RF electrodes induces a corresponding RF voltage on each corona ring through the intentionally small electrode--ring capacitance $C_{se}$; every stage couples to the same RF field, so the AC drive is effectively applied in parallel across all stages while the rectified DC contributions add in series along the column~\cite{cleland_industrial_2006,galloway_new_2004}. For insulation, PFSCR generators are almost universally housed in grounded pressure vessels filled with a high-dielectric-strength gas. The circuit topology adopted here is shown in Figure~\ref{fig:pfscr_topology}.

The PFSCR topology offers several practical advantages over conventional series-fed multipliers. First, the distributed RF drive reduces the effective source impedance at the high-voltage terminal and improves voltage regulation under beam load. Second, because the coupling between the RF electrodes and each rectifier stage is intentionally minimized, the total stored energy in the high-voltage system remains comparatively low ($\lesssim 100$~J) at full terminal voltage~\cite{galloway_new_2004-1}. This figure lies below the $\sim$195 J stored-energy bound empirically validated on the 300 kV prototype (Section~\ref{sec:lessons}), and consequently substantially limits the damage that can occur during internal discharges or vacuum sparks.

High-power classical PFSCR generators commonly employ industrial vacuum electron tubes (triodes), typically operating in Class C mode to drive the resonant tank~\cite{thompson_design_1969,thompson_high-power_1989,galloway_new_2004,matsuyama_improvement_2013}. The use of vacuum electron tubes drops the overall system efficiency to around 60\%~\cite{galloway_new_2004-1}. They also induce a 120 Hz terminal voltage ripple~\cite{thompson_multi-loop_1967} and have a limited lifespan, typically between 20,000 and 25,000 h~\cite{galloway_new_2004}. While more efficient generators based on solid-state oscillators have been designed, they have not yet been successfully implemented for voltages as high as 5 MV at power levels in the hundreds of kilowatts.

\subsubsection*{Circuit sizing}
The DC output voltage of a parallel-fed circuit under load comprises the no-load voltage and a load-dependent voltage drop~\cite{thompson_design_1969}:
\begin{equation}\label{eq:vout}
V_{0} = \frac{N V_{rf}}{K} - \frac{I(N-1)}{f C_{se} K},
\end{equation}
where $V_{rf}$ is the amplitude of the oscillator RF voltage, $N$ is the number of voltage-doubler stages, $I$ is the load current, $f$ is the operating frequency, $C_{se}$ is the coupling capacitance per stage, and $K$ is the coefficient of capacitive coupling,
\begin{equation}
K = 1 + 4\,\frac{C_{ac}}{C_{se}}.
\end{equation}
Here $C_{ac}$ is the shunt capacitance per stage. The peak-to-peak voltage ripple at the output is~\cite{thompson_design_1969}
\begin{equation}\label{eq:ripple}
V_{ripple} = \frac{I}{2 f C_{se}}.
\end{equation}
The design target is $V_{0}=5$~MV, $I=20$~mA, with a ripple factor below 0.5\%. Fixing two degrees of freedom from the oscillator ($V_{rf}=150$~kV and $f=250$~kHz) reduces the problem to finding $N$, $C_{se}$, and $K$. A parameterized geometry of the electrode column is developed in COMSOL Multiphysics, from which three capacitance components are computed: the capacitance between each RF electrode and the vessel ($C_a$), the capacitance between each set of corona rings and the adjacent RF electrode ($C_b$), and the capacitance between each set of corona rings and the neutral RF plane ($C_c$).

The choice of 250 kHz allows a significant reduction in voltage drop and ripple factor (a factor of $>2$ compared to the state of the art), at the cost of higher switching and diode dissipation and increased sensitivity to leakage inductance --- issues that are addressed in the thermal and oscillator designs below.

The configuration of the electrode column that satisfies the electrostatic stand-off requirements (Section~\ref{sec:hv_deliv}) is obtained at $N=120$. Table~\ref{tab:capacitances} summarizes the capacitance values for this configuration; $C_d$ represents the total junction capacitance of the diodes used in a single voltage-multiplier stage.

\begin{table}[t]
\caption{Capacitance components in the HV generator column.}
\label{tab:capacitances}
\centering
\begin{tabular}{@{}lc@{}}
\toprule
Parameter & Value\\
\midrule
$C_a$ & 147.6 pF\\
$C_b$ & 86.9 pF\\
$C_c$ & 61.2 pF\\
$C_d$ & 21.3 pF\\
\bottomrule
\end{tabular}
\end{table}

From Table~\ref{tab:capacitances}, the stage coupling capacitance is $C_{se}=2C_b/N=1.45$~pF and the shunt coupling capacitance per stage is $C_{ac}=(C_c+C_d)/(2N)=0.39$~pF, giving $K=2.07$. Using Eqs.~(\ref{eq:vout})--(\ref{eq:ripple}), the maximum terminal DC voltage is approximately 5.5~MV and the peak-to-peak voltage ripple is $\sim$27.6~kV (approximately 0.55\% of the terminal voltage). The ripple can be attenuated by an RF choke $L_i$ combined with the total capacitance from the terminal ($C_D$) and the plasma load ($C_L$):
\begin{equation}
V_{ripple,residual} \approx \frac{V_{ripple}}{\omega_s^2 L_i (C_D + C_L)}.
\end{equation}
With $C_D=26.0$~pF and $C_L=89.9$~pF, $L_i>3.8$~mH is required to reduce the ripple factor below 0.5\%.

\subsection{RF oscillator: CLLC resonant tank}\label{sec:cllc}
\begin{figure*}[t]
\centering
\resizebox{0.8\textwidth}{!}{%
\begin{circuitikz}[american currents, american voltages, line width=0.5pt, big/.style={circuitikz/blocks/scale=3},
long/.style={circuitikz/bipoles/twoportsplit/width=1.5}, 
longL/.style = {inductor, inductors/scale=1.5,
inductors/width=0.75, inductors/coils=10}, 
shortL/.style = {inductor, inductors/scale=1.5,
inductors/width=0.75, inductors/coils=3}, 
smallC/.style = {capacitor, capacitors/scale=0.5,
capacitors/width=0.2}]
\ctikzset{block lateral anchors pos=0.8}
\def\xacdc{0.5}
\def\xInvL{4.5}
\def\xInvR{6.5}
\def\xTank{8.5}
\def\xTr{14.50}
\def\xOut{18.0}
\def\yacdc{5}
\def\yP{9.0}
\def\yDp{9}
\def\yDn{3}
\def\yA{7}
\def\yB{5}
\def\yN{5}
\def\yC{3}
\def\yM{1.0}
\def\yTop{6}
\def\yRet{4.0}
\path (\xacdc,\yacdc) node[sacdcshape, big, circuitikz/block right anchors pos=0.2, label={[font=\footnotesize]center:$AC{-}DC$}](A){}
(3+\xacdc,\yacdc);
\node[nigfete, anchor=D, bodydiode] (Q1) at (\xInvL,8.25) {};
\node[nigfete, anchor=D, bodydiode] (Q2) at (\xInvL,2.75) {};
\draw (\xInvL,\yP) -- (Q1.C);
\draw (Q1.E) -- (Q2.C) coordinate (MID_L);
\draw (Q2.E) -- (\xInvL,\yM);
\draw (Q1.D) -- ++(0,0) node[left]{$Q_1$};
\draw (Q2.D) -- ++(0,0) node[left]{$Q_2$};
\draw (\xInvR,\yP) -- (\xInvL-1.75,\yP) -- (\xInvL-1.75,\yP-3.775) -- (A.right up);
\draw (\xInvR,\yM) -- (\xInvL-1.75,\yM) -- (\xInvL-1.75,\yM+3.775) -- (A.right down);
\draw (A.left up) -- ++(-1,0) to[sV] ++(0,-1.76) -- (A.left down);
\node[nigfete, anchor=D, bodydiode] (Q3) at (\xInvR,8.25) {};
\node[nigfete, anchor=D, bodydiode] (Q4) at (\xInvR,2.75) {};
\draw (\xInvR,\yP) -- (Q3.C);
\draw (Q3.E) -- (Q4.C) coordinate (MID_R);
\draw (Q4.E) -- (\xInvR,\yM);
\draw (Q3.D) -- ++(0,0) node[left]{$Q_3$};
\draw (Q4.D) -- ++(0,0) node[left]{$Q_4$};
\node[align=center] at (0.25,9.75) {AC-DC Converter};
\node[align=center] at (-2+\xInvR,9.75) {Full-bridge Inverter};
\node[align=center] at (11.75, 9.75) {Resonant Tank and Load};
\draw[dashed] (2.25, 9.75) -- (2.25,0.25);
\draw[dashed] (7.05, 9.75) -- (7.05,0.25);
\coordinate (DRV) at (MID_R);
\coordinate (RET) at (MID_L);
coordinate (\xTank-0.8,\yTop)  (DRVt);
\draw (\xInvL, \yTop-2) to[short,*-] (\xTank-0.8,\yRet) coordinate (RETr);
\draw (\xInvR,\yTop) to[short,*-] ++(0.75,0) to[L,l=$L_s$] ++(1,0) to[C,l=$C_s$] ++(1,0) coordinate (X);
\draw ((\xInvR+2.75,\yTop) to[short] ++(1,0);
\draw ((\xInvR+3.25,\yTop) to[C,l_=$C_r$] ++(0,-2);
\draw (\xInvR+3.6,\yTop) to[short] ++(1,0)
    to [shortL, name=l1, l_=$L_1$] ++(0,-2)
    to[short] ++(-4,0)
    (\xInvR+5.6,\yP) to[short] ++(0,-2) to[longL,mirror, name=l2_top] ++(0,-2) to[longL,mirror, name=l2_down] ++(0,-2) to[short] ++(0,-2)
    (\xInvR+5.6,\yP) to[short] ++(1.5,0) to[smallC, l=$C_a$] ++(0,-4) to[smallC, l=$C_a$] ++(0,-4)
   (\xInvR+5.6,\yM) to[short] (\xInvR+9.6,\yM)
    to[smallC, l=$C_b$] ++(0,2)
    to[short] ++(-1,0) to[smallC,l_=$C_c$] ++(0,2) to[smallC,l_=$C_c$] ++(0,2)
    (\xInvR+9.6, \yM+2) to[short] ++(1,0) to[smallC,l_=$C_d$] ++(0,2) to[smallC,l_=$C_d$] ++(0,2) to[short] ++(-2,0)
    (\xInvR+9.6, -2+\yP) to[smallC, l=$C_b$] ++(0,2)
    (\xInvR+5.6,\yP) to[short] ++(4.9,0)
    (\xInvR+9.6,\yM)  to[short] ++(0.9,0)
    (\xInvR+10.5,\yP) to[short] ++(2,0)
    to[R, l_=$R_L$] ++(0,-8)
    to[short] ++(-2,0)
    (\xInvR+5.6,-4.25+\yP) node[right, yshift=0.25cm] {$L_2$};
    \path (l1.ul dot) node[circ]{} (l2_top.ul dot) node[circ]{};
    \draw ([yshift=0.9cm]l1.north) 
    to[bend left=90] node[midway, above, yshift=0.2cm] {$M$} 
    ([xshift=-0.3cm, yshift=0.35cm]l2_top.west);
\end{circuitikz}%
}
\caption{Overall architecture of the RF oscillator with the PFSCR small-signal equivalent network.}
\label{fig:oscillator_overall}
\end{figure*}
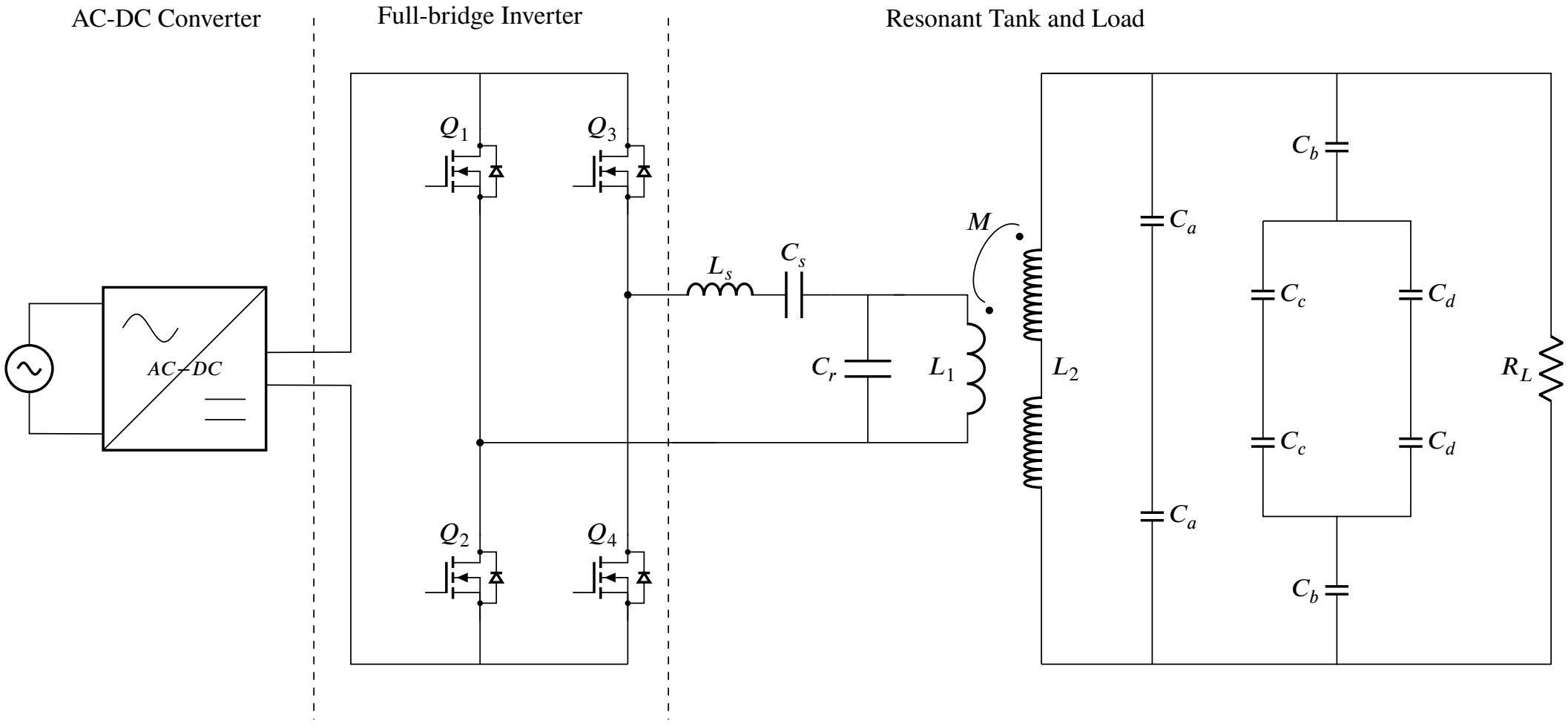

The RF oscillator adopted here is a CLLC resonant converter, which is a fully soft-switched topology with a wide voltage-gain range~\cite{zhang_introduction_2023}; it is widely used for interfacing DC buses in renewable generation, electric vehicles, and DC microgrids~\cite{zhao_design_2018,nhat_trung_tran_hybrid_2025}. The uni-directional CLLC variant has been demonstrated as a feasible RF oscillator in a 1.3 MV Dynamitron~\cite{gong_solid-state_2024}. The proposed oscillator is shown in Figure~\ref{fig:oscillator_overall} and comprises four stages: (1) a commercial, fast-response AC--DC converter providing the input DC bus; (2) a SiC-MOSFET full bridge ($Q_1$--$Q_4$) producing a high-frequency square wave; (3) a series LC branch ($L_s$, $C_s$) that filters higher-order harmonics; and (4) the resonant tank proper, consisting of an air-core step-up transformer and parallel resonant capacitors (discrete $C_r$ on the primary side, and the PFSCR capacitor network on the secondary).

\subsubsection*{CLLC resonant network}
\begin{figure}[H]
\centering
\resizebox{0.35\textwidth}{!}{%
\begin{circuitikz}[
longL/.style = {inductor, inductors/scale=1.5,
inductors/width=0.75, inductors/coils=10}, 
shortL/.style = {inductor, inductors/scale=1.5,
inductors/width=0.75, inductors/coils=3}]
\def\yTop{4}
\def\yM{1.0}
    \draw
    (-0.5,\yTop) to [open, v=$V_s$,o-o] (-0.5,0) 
    (-0.5,\yTop) to[short] ++(1.5,0)
    to[C, l_=$C_{r}$, mirror] ++(0,-\yTop)
    (1,\yTop) to[short] ++(1.5,0)
    to [shortL, name=l1, l_=$L_1$] ++(0,-\yTop)
    to[short] (-0.5,0)
    (4,\yTop) to[short] ++(0,-0.5) to[longL,mirror, name=l2_top] ++(0,-1.5) to[longL,mirror, name=l2_down] ++(0,-1.5) to[short] ++(0,-0.5)
    to[short] (5.5,0)
    to[C, l=$C_T$] ++(0,\yTop)
    (4,\yTop) to[short] ++(1.75,0)
    (5.75,\yTop) to[short] (7,\yTop)
    to[R, l_=$R_L$] (7,0)
    to[short] (5.5,0)
    (3.25,0.45*\yTop) node[right, yshift=0.25cm] {$L_2$}
    (7.5,\yTop) to [open, v=$V_{rf}$] (7.5,0); 
    \path (l1.ul dot) node[circ]{} (l2_top.ul dot) node[circ]{};
    \draw ([yshift=0.9cm]l1.north) 
    to[bend left=90] node[midway, above, yshift=0.75] {$M$} 
    ([xshift=-0.3cm, yshift=0.35cm]l2_top.west);
\end{circuitikz}}
\caption{Equivalent circuit of the CLLC resonant tank.}
\label{fig:tank_trans}
\end{figure}
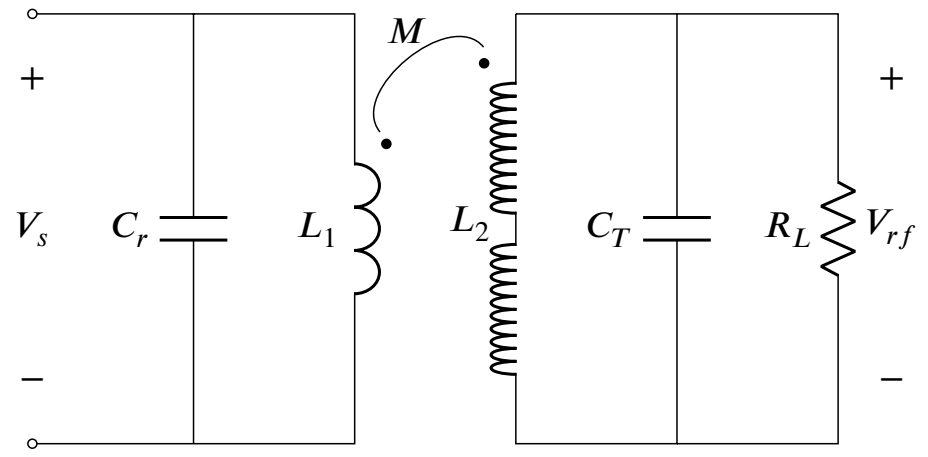

Figure~\ref {fig:tank_trans} shows the simplified resonant network following the LC filter. The secondary resonant capacitor $C_T$ is dominated by the equivalent capacitance of the PFSCR ladder, $0.5(C_a + C_b(C_c+C_d)/(C_b+C_c+C_d))$, with the parasitic capacitance of the secondary winding added. The transformer is designed so that the secondary leakage inductance satisfies $\omega_s^2 = 1/L_{\sigma 2} C_T$. Under this condition, the gain of the CLLC network is
\begin{equation}\label{eq:rf2sq}
\frac{V_{rf}}{V_s} = \omega R_L C_T \frac{M}{L_1}.
\end{equation}

For fixed frequency ($\omega=\omega_s$), load $R_L$, and PFSCR architecture, the transformer must be designed so that the ratio $M/L_1$ provides the required voltage boost. Reaching a 1 kV\textsubscript{pk} to 150 kV\textsubscript{pk} gain requires $M/L_1 > 10.6$. The primary-side current draw is
\begin{equation}\label{eq:is_big}
i_s = V_s \frac{L_2 \left(1 + \dfrac{j\omega C_T M^2}{L_1 L_2} R_L - \omega^2 L_{\sigma 1} C_r\right)}{j\omega L_1 L_{\sigma 2}}.
\end{equation}
Given $L_{\sigma 1}$, the primary resonant capacitor is sized to satisfy $\omega_s^2 = 1/L_{\sigma 1} C_r$, so the tank presents a purely resistive load to the inverter, which minimizes reactive-power circulation. Equation~(\ref{eq:is_big}) then simplifies to
\begin{equation}\label{eq:is_small}
i_s = V_s \frac{C_T M^2 R_L}{L_1^2 L_{\sigma 2}}.
\end{equation}

\subsection{Transformer design}\label{sec:transformer}
Operation near 250 kHz makes an air-core transformer the natural choice for the CLLC tank. Two constraints fix the design: $L_{\sigma 2} = 1/(\omega_s^2 C_T) = 4.2$~mH, and $M/L_1 \geq 10.8$. Winding insulation must withstand a peak voltage of 1 kV on the primary and 150 kV on the secondary. To mitigate the AC-resistance penalty from the skin effect, both windings use 42 AWG (American Wire Gauge) litz wire bundles with a strand diameter of 73.7~$\mu$m. Table~\ref{tb:winding} lists the per-winding parameters. The secondary wire is commercially available; the primary wire requires custom bundling of nine standard $6750\times$ 42 AWG litz strands.

\begin{table}[t]
\caption{Parameters of the selected litz-wire winding bundles.}
\label{tb:winding}
\centering
\resizebox{\columnwidth}{!}{%
\begin{tabular}{@{}lccccc@{}}
\toprule
Winding & \shortstack{Ampacity \\ (A)} & \shortstack{No. of 42 \\ AWG strands} & OD (mm) & \shortstack{DC resistance \\ ($\Omega$/km)} & $R_{AC}/R_{DC}$ \\
\midrule
Primary   & 500 & 60,750 & 110.2 & 0.11 & 1.50 \\
Secondary & 30  & 4,200  & 7.3   & 1.53 & 1.55 \\
\bottomrule
\end{tabular}}
\end{table}

\begin{figure}[H]
    \centering
    \includegraphics[width=0.4\textwidth]{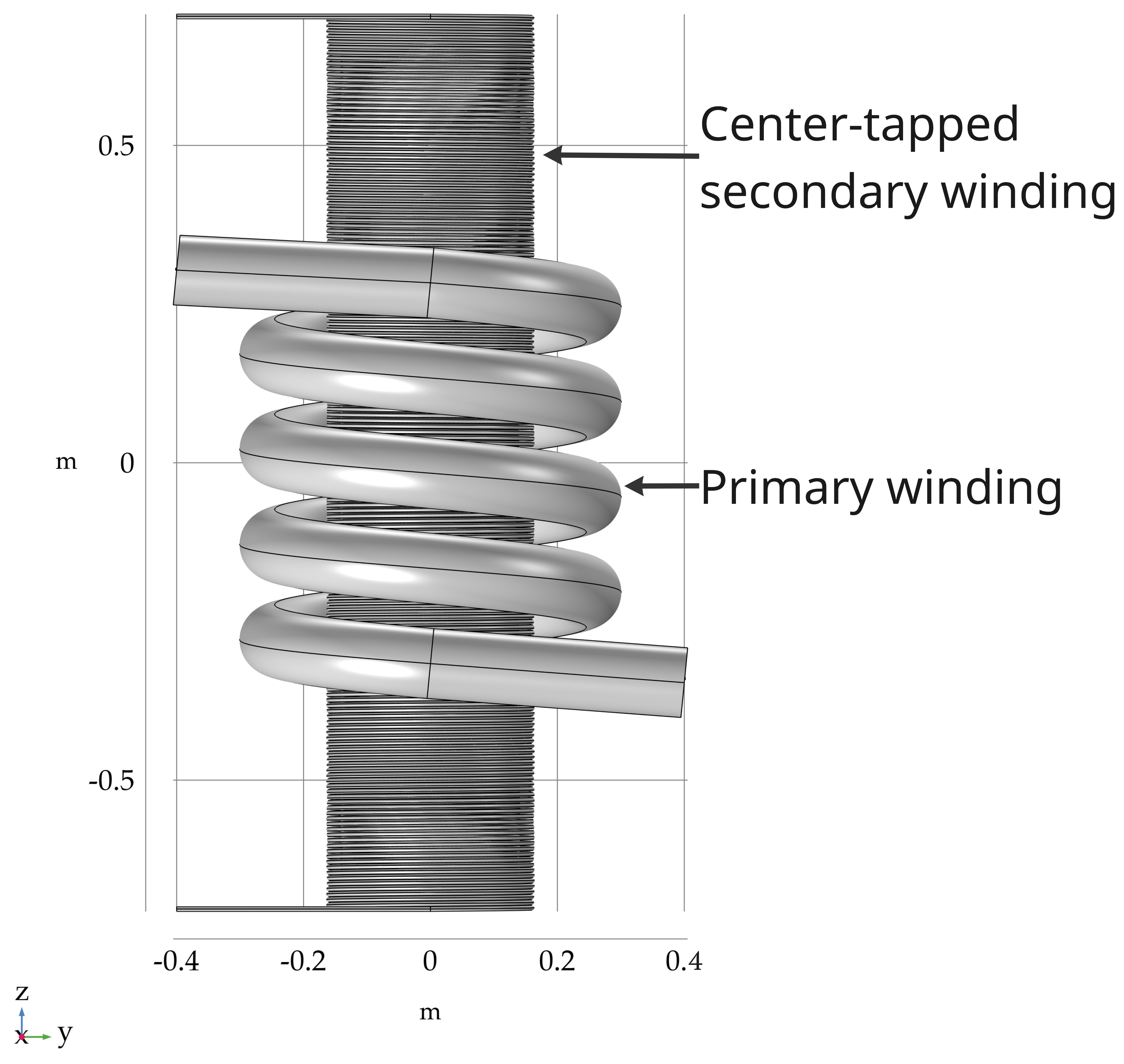}
    \caption{Finite-element model of the RF transformer.}
    \label{fig:transformer}
\end{figure}

The transformer is shown in Figure~\ref{fig:transformer}; outer diameter is 50~cm and height is 150~cm. The structural support is silicon nitride (Si\textsubscript{3}N\textsubscript{4}), chosen for its combination of thermal conductivity (28 W/m.K), dielectric strength (19 MV/m), and mechanical strength. The secondary winding has a mean radius of 15~cm and a length of 140~cm, sectioned into two halves with the center grounded (center-tapped). Its 320 turns are configured in two sections, each with two radial layers and 80 axial layers. Turn-to-turn polyimide insulation is $\sim$1.5~mm thick, providing substantial turn-to-turn hold-off. The primary winding consists of 4 turns at a mean radius of 25~cm. The coupling factor is approximately 49\%, which is acceptable given the air core and the very small primary turn count. Table~\ref{tab:transformer} lists the 3D-simulated transformer parameters.

The primary-to-secondary insulation must withstand a peak voltage of 150~kV. A 6~cm radial gap between the two windings, combined with the dielectric strength of the pressurized insulating gas, readily meets this requirement.

The simulated secondary leakage inductance matches the target $L_{\sigma 2} = 4.2$~mH within 0.1\%. With $L_{\sigma 1} = 2.93~\mu$H, the primary resonant capacitor is $C_r = 1/(\omega_s^2 L_{\sigma 1}) = 138.55$~nF.

\begin{table}[t]
\caption{Transformer parameters from 3D simulation.}
\label{tab:transformer}
\centering
\begin{tabular}{@{}lc@{}}
\toprule
Parameter & Value\\
\midrule
Primary inductance    & 3.84 $\mu$H\\
Secondary inductance  & 5.51 mH\\
Mutual inductance     & 71.04 $\mu$H\\
Primary AC resistance & 20.37 $\mu\Omega$\\
Secondary AC resistance & 128.66 m$\Omega$\\
\bottomrule
\end{tabular}
\end{table}

\subsection{Efficiency estimate}\label{sec:efficiency}
The power losses from the major contributors are summarized in Table~\ref{tab:loss_budget}. Transformer ohmic losses are computed from the winding currents under resonance:
\begin{subequations}
\begin{align}
i_1 &= V_s \frac{j\omega L_1 L_2 L_{\sigma 2} - M^2 R_L}{(j\omega L_1 L_{\sigma 2})^2} \label{eq:1a}\\
i_2 &= V_s \frac{M L_1 R_L - j\omega M L_1 L_{\sigma 2}}{(j\omega L_1 L_{\sigma 2})^2} \label{eq:1b}
\end{align}
\end{subequations}
For $V_s=875$~V (which yields 5~MV at the output at full load), the primary and secondary windings dissipate 5.1~W and 115.8~W respectively.

The switch, diode, and tank losses come from LTSpice circuit simulations at full load (5 MV, 20 mA). The inverter uses Wolfspeed CAB500M17HM3 half-bridge module that uses SiC MOSFETs. The series filter inductor $L_s$ is 10 turns of the same wire as the transformer primary, with a coil diameter of 15~cm and a length of 125~cm, giving 1.03~$\mu$H and 0.54~m$\Omega$. The series resonant capacitor $C_s = 390$~nF is a CalRamic CR09 ultra-low-loss pulsed capacitor (0.2\% dissipation factor). Each rectifier stage uses 15 Voltage Multipliers Inc.\ X100UFG diodes in series.

With total losses below 7\% of the output power, the 1 kV DC to 5 MV DC conversion stage can theoretically achieve $>$93\% efficiency at full load. Assuming a commercial AC--DC front-end (e.g.\ Magna-Power MT-150kW) at 92\%, the total system efficiency is $\sim$85\%, a significant improvement on the $\sim$60\% typical of triode-driven Dynamitrons.

\begin{table}[t]
\caption{Major power-dissipation components at full load.}
\label{tab:loss_budget}
\centering
\begin{tabular}{@{}lc@{}}
\toprule
Component & Power loss (W)\\
\midrule
RF transformer   & 121 \\
Diodes           & 3,000 \\
Inverter MOSFETs & 2,980 \\
Series LC filter & 182 \\
\textbf{Total}   & \textbf{6,283} \\
\bottomrule
\end{tabular}
\end{table}

\subsection{Passive-component insulation and sourcing}\label{sec:components}
Table~\ref{tab:components} summarizes the voltage, current, and energy ratings required of each major passive and active component of the HV generation chain, along with its commercial availability status. Every component has been identified as either off-the-shelf or a custom variant that can be produced from a commercial base part through bundling, stacking, or repotting, without requiring the development of new technology.

\begin{table*}[t]
\caption{Major components, insulation requirements, and commercial availability.}
\label{tab:components}
\centering
\renewcommand{\arraystretch}{1.15}
\begin{tabular}{@{}p{4.0cm}p{4.5cm}p{3.2cm}p{4.0cm}@{}}
\toprule
Component & Peak rating & Source & Notes \\
\midrule
AC--DC front end    & 150 kW, 1 kV DC   & Commercial (Magna-Power MT-150kW) & 92\% efficiency \\
Inverter Half-bridge module    & 1.7 kV, 500 A each & Commercial (Wolfspeed CAB500M17HM3) & Soft-switched SiC \\
Series LC filter $L_s$ & $\sim$1 kV\textsubscript{pk}, 1~$\mu$H & Custom (litz-wound) & Same wire as transformer primary \\
Series resonant $C_s$ & 2 kV, 390 nF & Commercial (CalRamic CR09) & 0.2\% dissipation factor \\
Air-core transformer & 150 kV\textsubscript{pk} secondary, 500 A primary & Custom & Custom 60,750-strand litz primary \\
Rectifier diodes (each of $120\times 15$) & 100 kV, 50 mA & Commercial (VMI X100UFG) & Series-stacked in each doubler \\
Parallel-fed coupling $C_{se}$ & 150 kV\textsubscript{pk}, 1.45 pF & Integrated into corona-ring geometry & No discrete component required \\
RF choke $L_i$ & 5 MV standoff, 3.8 mH, 20 mA & Custom & Located inside the pressurized vessel \\
Bushing insulators & 100 kV/spacer, 50 spacers & Commercial blanks, custom machined & 97\% Al\textsubscript{2}O\textsubscript{3}, brazeable to tungsten \\
Insulating gas      & 1 MPa, $\sim$50 MV/m strength & Commercial blend & C\textsubscript{5}F\textsubscript{10}O / CO\textsubscript{2} / O\textsubscript{2} at 2.75/91.75/5.5\% \\
\bottomrule
\end{tabular}
\end{table*}

\subsection{Sensing and regulation}\label{sec:sensing}
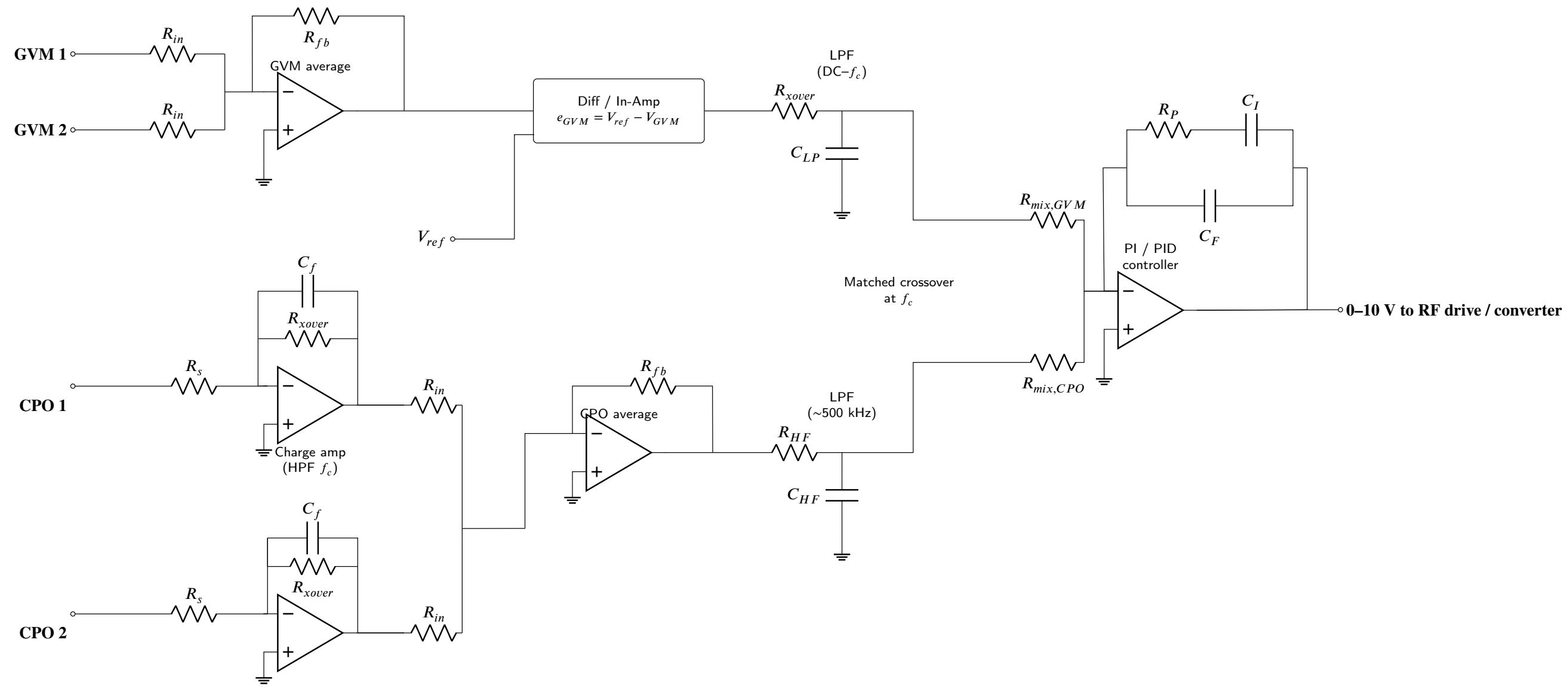
\begin{figure*}[t]
\centering
\resizebox{\textwidth}{!}{%
\ctikzset{bipoles/length=1.15cm}
\begin{circuitikz}[american, scale=0.9, transform shape]
    \tikzstyle{protext} = [font=\footnotesize\sffamily, align=center]
    \tikzstyle{block}   = [draw, rounded corners=2pt, minimum width=3.6cm,
                           minimum height=1.35cm, font=\footnotesize\sffamily, align=center]
    \def\xIn{-1.0}
    \def\xSum{4.0}
    \def\xBlk{10.5}
    \def\xFlt{15.2}
    \def\xCtrl{22.2}
    \def\xOut{25.2}
    \def\yGVMa{11.2}
    \def\yGVMb{9.6}
    \def\yVref{7.3}
    \def\yCPOa{3.8}
    \def\yCPOb{-1}
    \def\yCtrl{6.2}
    \def\yMixG{7.7}
    \def\yMixC{4.7}
    \coordinate (GVM1_IN) at (\xIn,\yGVMa);
    \coordinate (GVM2_IN) at (\xIn,\yGVMb);
    \coordinate (VREF_IN) at (8+\xIn,\yVref);
    \coordinate (CPO1_IN) at (\xIn,\yCPOa);
    \coordinate (CPO2_IN) at (\xIn,\yCPOb);
    \node[left, font=\bfseries] at (GVM1_IN) {GVM 1};
    \node[left, font=\bfseries] at (GVM2_IN) {GVM 2};
    \node[left, font=\bfseries] at (VREF_IN) {$V_{{ref}}$};
    \node[left, font=\bfseries] at (CPO1_IN) {CPO 1};
    \node[left, font=\bfseries] at (CPO2_IN) {CPO 2};
    \node[op amp] (op_gvm_sum) at (\xSum,10.0) {};
    \draw (op_gvm_sum.+) -- ++(0,-0.7) node[ground]{};
    \coordinate (GVM_SUM_NODE) at (\xSum-1.8,10.4);
    \draw (GVM1_IN) to[short, o-] ++(1.0,0)
          to[R, l=$R_{in}$] (\xSum-1.8,\yGVMa);
    \draw (\xSum-1.8,\yGVMa) -- (\xSum-1.8,10.0);
    \draw (GVM2_IN) to[short, o-] ++(1.0,0)
          to[R, l=$R_{in}$] (\xSum-1.8,\yGVMb);
    \draw (\xSum-1.8,\yGVMb) -- (\xSum-1.8,10.0);
    \draw (GVM_SUM_NODE) -- (op_gvm_sum.-);
    \draw (op_gvm_sum.out) -- ++(1,0) coordinate(gvm_sum_out);
    \draw (gvm_sum_out) -- ++(0,2) -- ++(-0.5,0) to[R, l=$R_{fb}$]  ++(-2.7,0) -- (\xSum-1.22,10.40);
    \node[protext, above=0.675cm] at (op_gvm_sum) {GVM average};
    \node[block] (blk_inamp) at (\xBlk,10.0)
      {Diff / In-Amp\\$e_{{GVM}}=V_{{ref}}-V_{{GVM}}$};
    \draw (gvm_sum_out) -- (\xBlk-2.2,10.0) -- (blk_inamp.west);
    \coordinate (VREF_TAP) at (\xBlk-2.2,\yVref);
    \draw (VREF_IN) to[short, o-] ++(1.0,0) -- (VREF_TAP);
    \draw (VREF_TAP) -- (\xBlk-2.2,9.5) -- (\xBlk-2.0,9.5) -- ++(0.215,0);
    \coordinate (GVM_LPF_RL) at (\xFlt-2.0,10.0);
    \coordinate (GVM_LPF_N)  at (\xFlt,10.0);
    \coordinate (GVM_LPF_OUT) at (\xFlt+1.5,10.0);
    \draw (blk_inamp.east) -- (GVM_LPF_RL)
          to[R, l=$R_{xover}$] (GVM_LPF_N) -- (GVM_LPF_OUT);
    \draw (GVM_LPF_N) to[C, l_=$C_{LP}$] (\xFlt,8.2) node[ground]{};
    \node[protext, above=0.55cm] at (GVM_LPF_N) {LPF\\(DC--$f_c$)};
    \node[op amp] (op_cpo1) at (\xSum,3.8) {};
    \draw (op_cpo1.+) -- ++(0,-0.2) node[ground]{};
    \draw (CPO1_IN)++(0,0.4) to[short, o-] ++(1.0,0)
          to[R, l=$R_s$] (\xSum-0.9,\yCPOa+0.4);
    \draw (\xSum-0.9,\yCPOa+0.4) -- (\xSum-0.9,4.2) -- (op_cpo1.-);
    \draw (op_cpo1.out) -- ++(1.0,0) coordinate(cpo1_out);
    \coordinate (cpo1_fb_top) at (\xSum-1.1,5.2);
    \draw (op_cpo1.-) -- (\xSum-1.1,4.2) -- (\xSum-1.1,5.2);
    \draw (cpo1_fb_top) -- ++(0,1) to[C, l=$C_f$] (\xSum+1.0,6.2) -- (\xSum+1.0,3.8) -- (cpo1_out);
    \draw (cpo1_fb_top) to[R, l=$R_{xover}$] (\xSum+1.0,5.2);
    \node[protext, below=0.75cm] at (op_cpo1) {Charge amp\\(HPF $f_c$)};
    \node[op amp] (op_cpo2) at (\xSum,-1) {};
    \draw (op_cpo2.+) -- ++(0,-0.2) node[ground]{};
    \draw (CPO2_IN)++(0,0.4) to[short, o-] ++(1.0,0)
          to[R, l=$R_s$] (\xSum-0.9,\yCPOb+0.4);
    \draw (op_cpo2.out) -- ++(1.0,0) coordinate(cpo2_out);
    \coordinate (cpo2_fb_top) at (\xSum-0.9,1);
    \draw (\xSum-0.9,-0.6) -- (\xSum-0.9,1);
    \draw (cpo2_fb_top) to[C, l=$C_f$] (\xSum+1.0,1) --  (\xSum+1.0,-1) -- (cpo2_out);
    \draw (cpo2_fb_top)-- ++(0,-0.6) to[R, l_=$R_{xover}$] (\xSum+1.0,0.4);
    \node[op amp] (op_cpo_sum) at (\xBlk,2.8) {};
    \draw (op_cpo_sum.+) -- ++(0,-0.2) node[ground]{};
    \node[protext, above=0.55cm] at (op_cpo_sum) {CPO average};
    \coordinate (CPO_SUM_NODE) at (\xBlk-1.8,2.8);
    \draw (cpo1_out) -- (\xBlk-4.5,3.8)
          to[R, l=$R_{in}$] (\xBlk-3.3,3.8);
    \draw (\xBlk-3.3,3.8) -- (\xBlk-3.3,1);
    \draw (cpo2_out) -- (\xBlk-4.5,-1) to[R, l=$R_{in}$] (\xBlk-3.3,-1);
    \draw (\xBlk-3.3,-1) -- (\xBlk-3.3,1);
    \draw (op_cpo_sum.out) -- ++(1.0,0) coordinate(cpo_sum_out);
    \draw (op_cpo_sum.-) -- ++(0,1) -- ++(0.45,0) to[R, l=$R_{fb}$] ++(2.5,0) -- (cpo_sum_out);
     \draw (op_cpo_sum.-) -- ++(-1,0) -- ++(0,-2) --  (\xBlk-3.3,1.2);
    \coordinate (CPO_HF_RL)  at (\xFlt-2.0,2.8);
    \coordinate (CPO_HF_N)   at (\xFlt,2.8);
    \coordinate (CPO_HF_OUT) at (\xFlt+1.5,2.8);
    \draw (cpo_sum_out) -- (CPO_HF_RL)
          to[R, l=$R_{HF}$] (CPO_HF_N) -- (CPO_HF_OUT);
    \draw (CPO_HF_N) to[C, l_=$C_{HF}$] (\xFlt,1.0) node[ground]{};
    \node[protext, above=0.55cm] at (CPO_HF_N) {LPF\\($\sim$500 kHz)};
    \node[op amp] (op_ctrl) at (-0.5+\xCtrl,-0.4+\yCtrl) {};
    \draw (op_ctrl.+) -- ++(0,-0.7) node[ground]{};
    \node[protext, above=0.75cm] at (op_ctrl) {PI / PID\\controller};
    \coordinate (CTRL_IN) at (\xCtrl-1.2,\yCtrl);
    \coordinate (GVM_LANE_L) at (\xFlt+1.5,\yMixG);
    \coordinate (GVM_LANE_R) at (\xCtrl-3.2,\yMixG);
    \draw (GVM_LPF_OUT) -- (\xFlt+1.5,10.0) -- (\xFlt+1.5,\yMixG) -- (GVM_LANE_R)
          to[R, l=$R_{mix,GVM}$] (\xCtrl-1.9,\yMixG) -- (\xCtrl-1.9,\yCtrl) -- (CTRL_IN) -- (op_ctrl.-);
    \coordinate (CPO_LANE_R) at (\xCtrl-3.2,\yMixC);
    \draw (CPO_HF_OUT) -- (\xFlt+1.5,2.8) -- (\xFlt+1.5,\yMixC) -- (CPO_LANE_R)
          to[R, l_=$R_{mix,CPO}$] (\xCtrl-1.9,\yMixC) -- (\xCtrl-1.9,\yCtrl) -- (CTRL_IN);
    \node[protext] at (\xCtrl-5.8,6.2) {Matched crossover\\at $f_c$};
    \coordinate (FB_UP) at (\xCtrl-1.5,\yCtrl+2.6);
    \coordinate (FB_SPLIT) at (\xCtrl-1,\yCtrl+2.6);
    \coordinate (FB_JOIN) at (\xCtrl+4.0,\yCtrl+2.6);
    \draw (op_ctrl.-) -- (FB_UP) -- (FB_SPLIT);
    \coordinate (FB_TOP_Y) at (\xCtrl-1,\yCtrl+3.4);
    \draw (FB_SPLIT) -- (\xCtrl-1,\yCtrl+3.4)
          to[R, l=$R_P$] (\xCtrl+0.75,\yCtrl+3.4)
          to[C, l=$C_I$] (\xCtrl+2.5,\yCtrl+3.4)
          -- (\xCtrl+2.5,\yCtrl+2.6);
    \draw (FB_SPLIT) -- (\xCtrl-1,\yCtrl+1.8)
          to[C, l_=$C_F$] (\xCtrl+2.5,\yCtrl+1.8)
          -- (\xCtrl+2.5,\yCtrl+2.6);
    \draw (\xCtrl+2.5,\yCtrl+2.6) -- (\xCtrl+2.8,\yCtrl+2.6) -- (\xCtrl+2.80,\yCtrl-0.39) -- (op_ctrl.out);
    \draw (op_ctrl.out) -- (\xOut+0.5,\yCtrl-0.39) to[short, -o]
          ++(0.0,0.0) node[right, font=\bfseries] {0--10 V to RF drive / converter};
\end{circuitikz}
}
\caption{Low-voltage regulation loop combining a slow, absolute GVM measurement with a fast capacitive pick-off (CPO) path.}
\label{fig:mv_reg_control}
\end{figure*}

Terminal-voltage measurement is a key signal in the system, both for Orbitron operation and as the primary input to the closed-loop regulator that stabilizes the output voltage. The conceptual design is shown in Figure~\ref{fig:mv_reg_control}.

At the multi-megavolt level, conventional resistive or capacitive dividers become physically large and increasingly sensitive to surface leakage, corona, and flashover constraints. Generating voltmeters (GVMs) are used here to measure the terminal DC voltage. A grounded rotor with cut-out sectors rotates in front of a segmented stator, periodically modulating the terminal-to-stator capacitance and generating a current proportional to $V_0\, dC_{GVM}(t)/dt$. After transimpedance conversion and carrier filtering, the low-voltage output is proportional to the terminal voltage. The useful bandwidth of a GVM channel is limited to a few Hz by the rotor speed and carrier filtering, which introduces phase lag that constrains the control-loop crossover frequency.

To capture fast disturbances that the GVM cannot follow, capacitive pick-off (CPO) plates are used. Each CPO plate is a metal electrode mounted on --- and insulated from --- the tank wall and facing the HV terminal, forming a small capacitance $C_{plate}$. The plate current is $C_{plate}\, dV_0/dt$, and a charge amplifier (transimpedance stage) with feedback capacitor $C_f$ and bleed resistor $R_f$ produces an output approximately proportional to the AC component of $V_0$ above $f_c = 1/(2\pi R_f C_f)$, while rolling off toward DC to avoid drift and saturation. An additional high-frequency low-pass stage (typically a few hundred kilohertz) suppresses direct RF pickup from the 250 kHz oscillator field.

Two GVM heads mounted on opposing sides of the terminal are averaged to reduce sensitivity to lateral motion and field asymmetry. The averaged GVM signal is compared to the setpoint $V_{ref}$ in a differential (instrumentation) amplifier to generate a slow but accurate error signal, which is low-pass filtered (DC to $f_c$). In parallel, two CPO plates are processed with charge amplifiers, averaged, and high-pass filtered above $f_c$, with an additional high-frequency low-pass to band-limit the fast path. The low-pass GVM error and the band-limited CPO correction are then combined using matched crossover filters so that control authority hands off smoothly between channels around $f_c$ without introducing gain peaking. The consolidated error drives a PI controller (with an optional strongly filtered lead term for phase margin). The controller output is limited to a 0--10 V command that interfaces with the PFSCR RF drive/converter control input, corresponding to 0--5 MV at the terminal.

\section{HV Distribution and Delivery}\label{sec:hv_deliv}
\begin{figure*}[b]
    \centering
    \includegraphics[width=0.99\textwidth, angle=0]{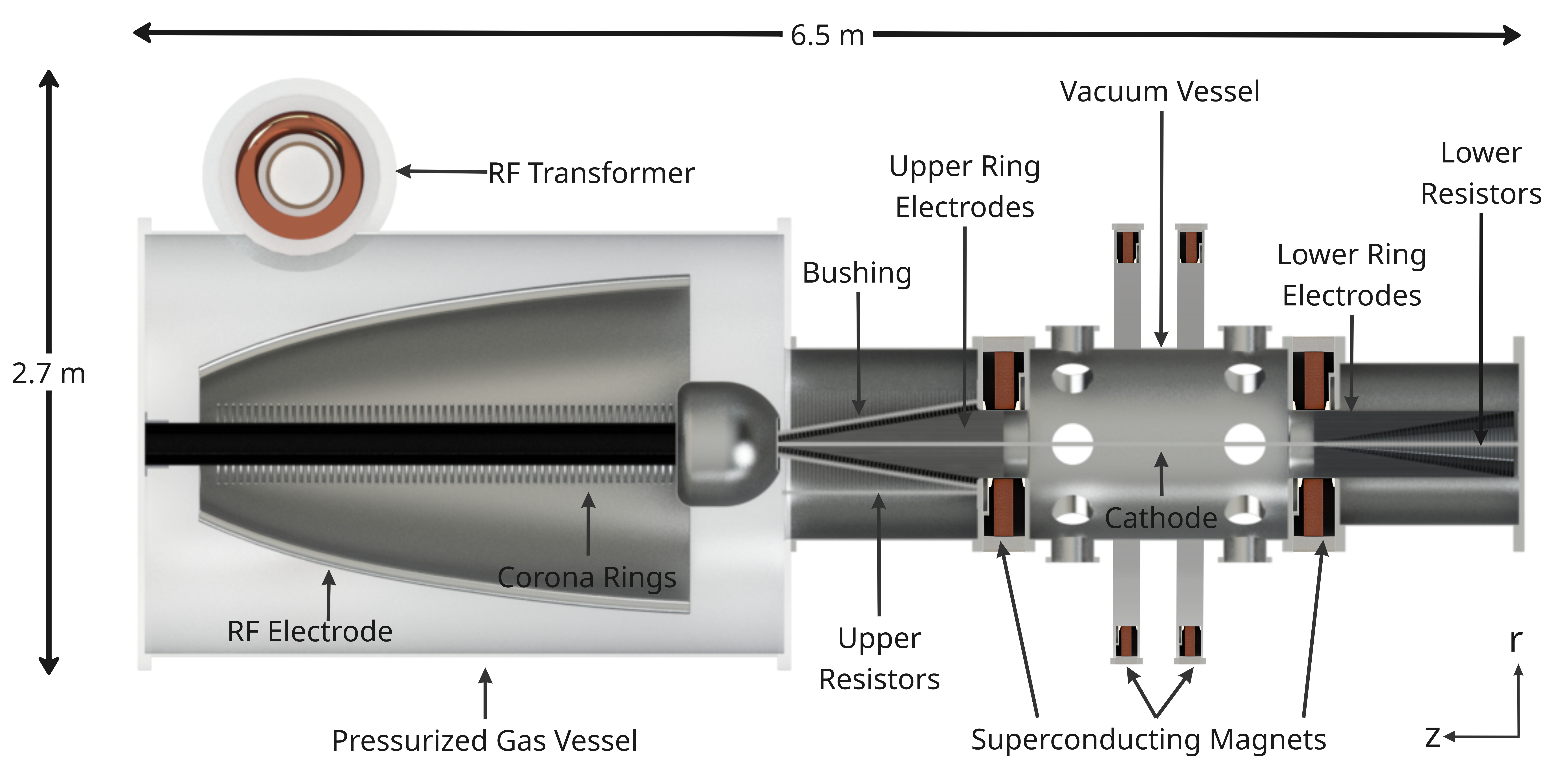}
    \caption{CAD model of the 5 MV Orbitron system, including the HV generator, the bushing, and the voltage-distribution network.}
    \label{fig:whole_thing}
\end{figure*}

The PFSCR generates terminal voltage, which is transmitted to the Orbitron's central cathode and distributed among 49 additional electrodes in the end electrode system via resistive division, as shown in Figure~\ref{fig:whole_thing}. Voltage generation and resistive division occur inside the pressurized gas medium; the vacuum bushing serves as the critical interface between that medium and the vacuum chamber where the fusion reactions take place.

\subsection{Pressurized gas vessel and insulating medium}\label{sec:gas_vessel}
Pressurized gas is used both for electrical insulation in a compact footprint and for thermal management: the heat produced by RF losses in the oscillator is removed by circulating the insulating gas through heat exchangers outside the tank (not shown in Figure~\ref{fig:whole_thing}). A predominant choice of insulating gas for HV systems in the MV range has been SF\textsubscript{6} because of its excellent insulating properties and chemical inertness~\cite{christophorou_sfsub_1995}. For environmental reasons, this design uses perfluoroketone C\textsubscript{5}F\textsubscript{10}O, which has a global-warming potential (GWP) below 1 compared with 23,500 for SF\textsubscript{6}~\cite{zhang_fundamental_2020}. While C\textsubscript{5}F\textsubscript{10}O has a critical dielectric strength twice that of SF\textsubscript{6}, it has a boiling temperature of $\sim$26.9$^\circ$C and its vapor pressure at 25$^\circ$C is only 93.8 kPa~\cite{simka_dielectric_2015}, so it must be diluted with a carrier gas. A C\textsubscript{5}F\textsubscript{10}O / CO\textsubscript{2} / O\textsubscript{2} mixture at 1 MPa with a ratio of 2.75 / 91.75 / 5.5\% exhibits approximately the same dielectric strength as SF\textsubscript{6} at 0.6 MPa~\cite{stoller_mixtures_2017}, which in a uniform field yields a dielectric strength of 50--60 MV/m \cite{kuffel_high_2000}.

\begin{figure}[h]
     \centering
     \begin{subfigure}[b]{0.48\textwidth}
         \centering
         \includegraphics[width=0.9\textwidth]{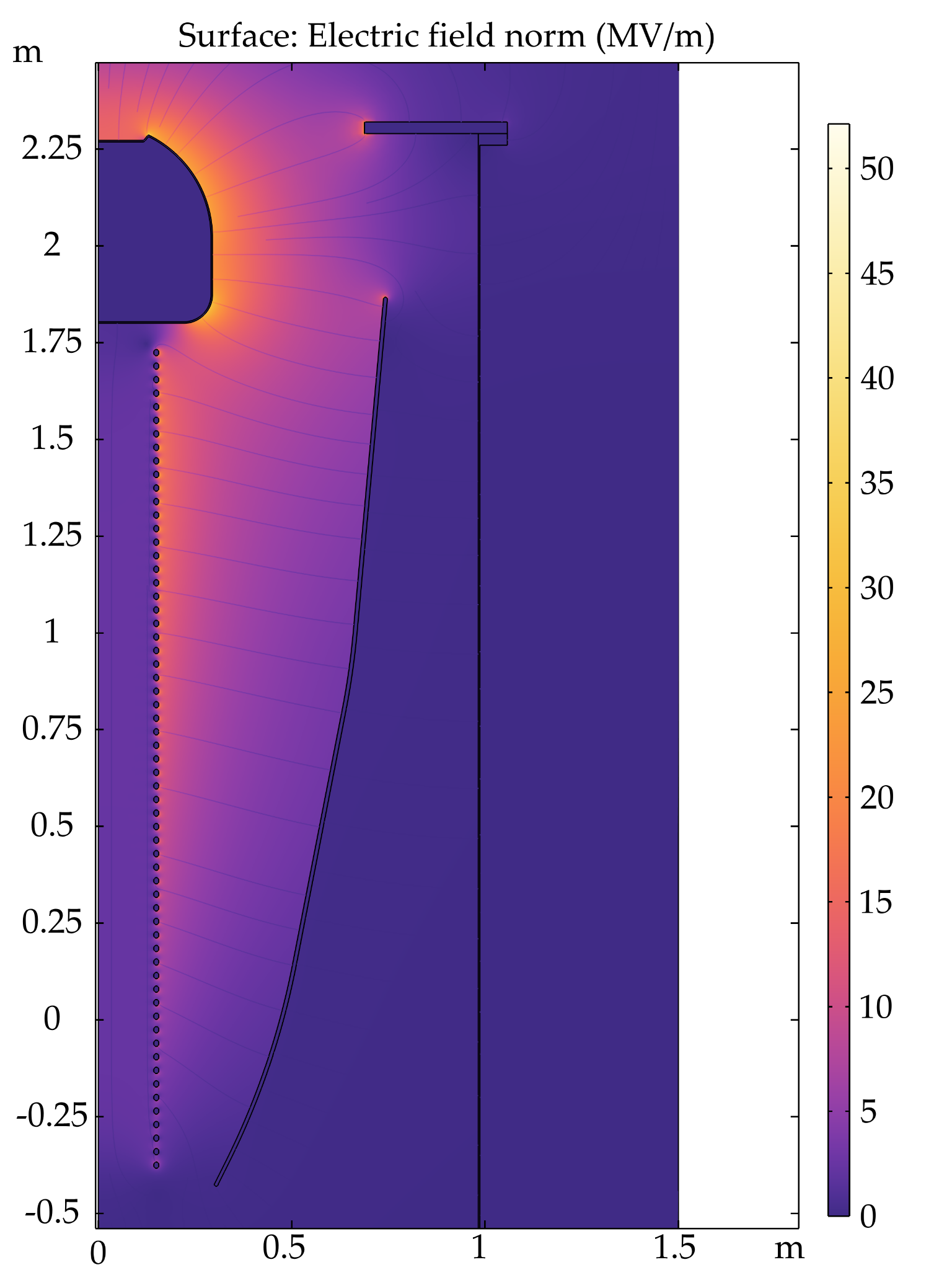}
         \caption{Side view.}
         \label{fig:dynam_side}
     \end{subfigure}
     \begin{subfigure}[b]{0.48\textwidth}
         \centering
         \includegraphics[width=\textwidth]{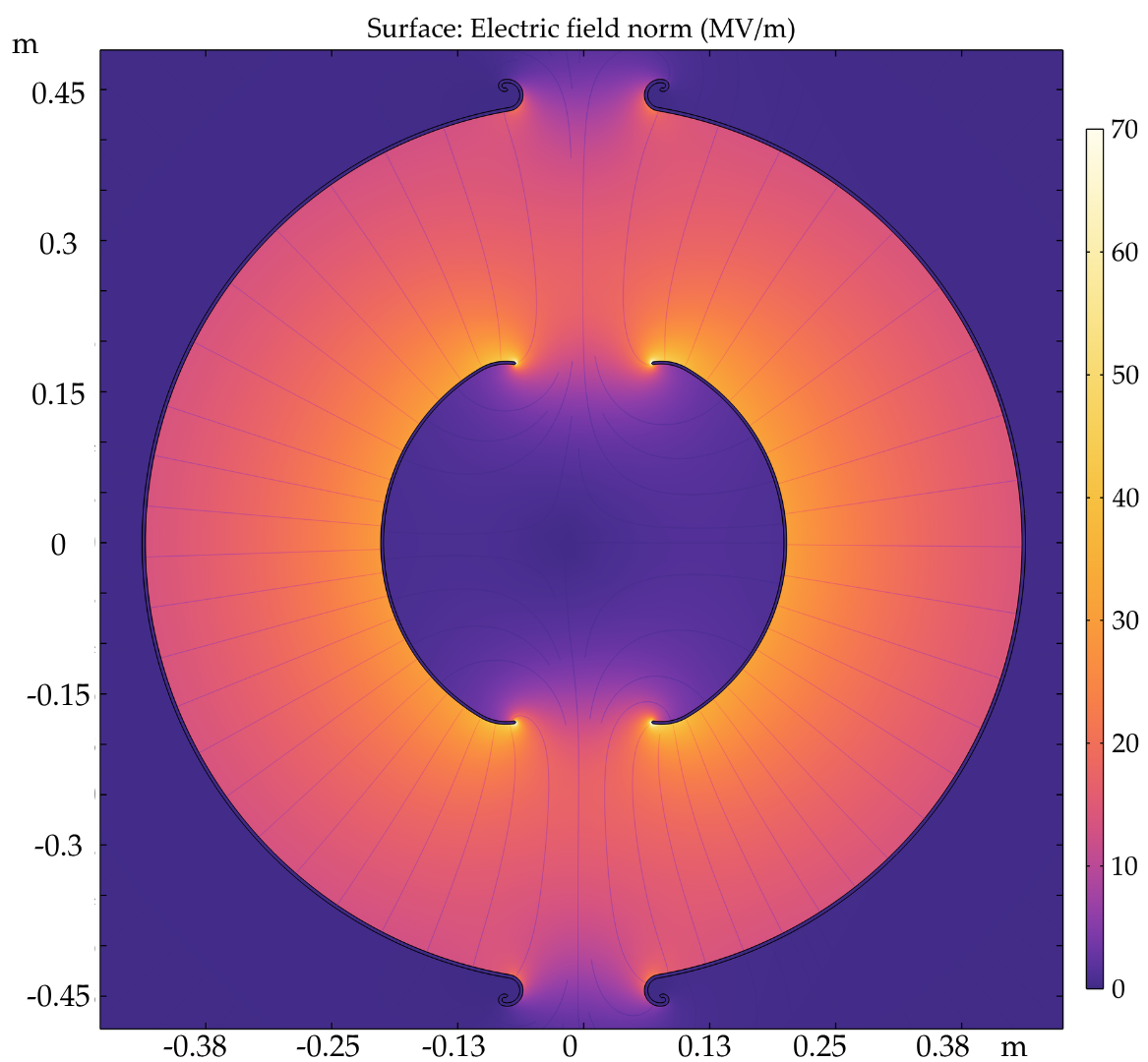}
         \caption{Top view.}
         \label{fig:dynam_top}
     \end{subfigure}
     \caption{Electric-field distribution in the PFSCR column at nominal 5 MV terminal voltage.}
     \label{fig:dynamitron_efield}
\end{figure}

The pressurized vessel, including the electrodes described above, is simulated in COMSOL Multiphysics using the steady-state values for each electrode at the nominal 5 MV terminal voltage. The resulting electric-field distributions from side and top views are plotted in Figure~\ref{fig:dynamitron_efield}. The corona rings are subject to two primary sources of electrical stress: (1) the 85~kV inter-ring potential difference, and (2) proximity to the RF electrode, which has the highest potential difference relative to the top corona rings. To mitigate the second, the RF electrode is curved and angled to minimize non-uniformity in the electric field. Under these conditions, the average electric field between consecutive corona rings is 11.9 MV/m, while the maximum surface field reaches 30.3 MV/m at the points where the top corona rings face the RF electrode. The average field between the top corona ring and the RF electrode is 8.7 MV/m.

\paragraph{Top-of-column field concentration}
The field distribution in Figure~\ref{fig:dynamitron_efield} shows a clear concentration near the HV generator terminal in the upper 1.8-2~m of the column, driven by the increasing potential difference between the RF electrodes and the grounded vessel wall. Three design measures manage this concentration. First, the corona rings' minor radii are sized to prevent corona inception in the topmost ring, resulting in a peak field of 30.3 MV/m at their surfaces. Second, the RF electrodes' curvature near the top of the column follows a quasi-Rogowski profile, keeping the ring-to-RF-electrode peak field below the gas mixture's streamer-inception field by at least a factor of 2. Third, the terminal dome itself extends upward with an oversized toroidal shield, moving the highest-field location away from any triple-junction or sealing feature. All three choices directly inherit the Hammerhead design philosophy of trapping field-emission-prone points inside equipotential wells (Section~\ref{sec:lessons}).

In addition to the Orbitron cathode, which connects directly to the PFSCR terminal, 49 additional electrodes lie between the cathode and the anode (vacuum vessel). Voltage is divided uniformly between them by a resistor string (one large resistor between each adjacent electrode pair). These electrodes serve two purposes: (1) they grade the voltage across the bushing, reducing the potential difference across individual insulating stages and mitigating the total-voltage effect along the graded electrode structure; and (2) they act as an end-plugging system that imposes a tailored potential profile on the plasma. In the pressurized-gas region, the horizontal segments of these electrodes are 0.75 mm thick and spaced 18 mm apart, giving an inter-electrode field of 5.5 MV/m. Their extension into the vacuum is carried through the vacuum bushing, described next.

\subsection{Vacuum bushing}\label{sec:bushing}
\begin{figure}[h]
     \centering
     \begin{subfigure}[b]{0.23\textwidth}
         \centering
         \includegraphics[width=\textwidth]{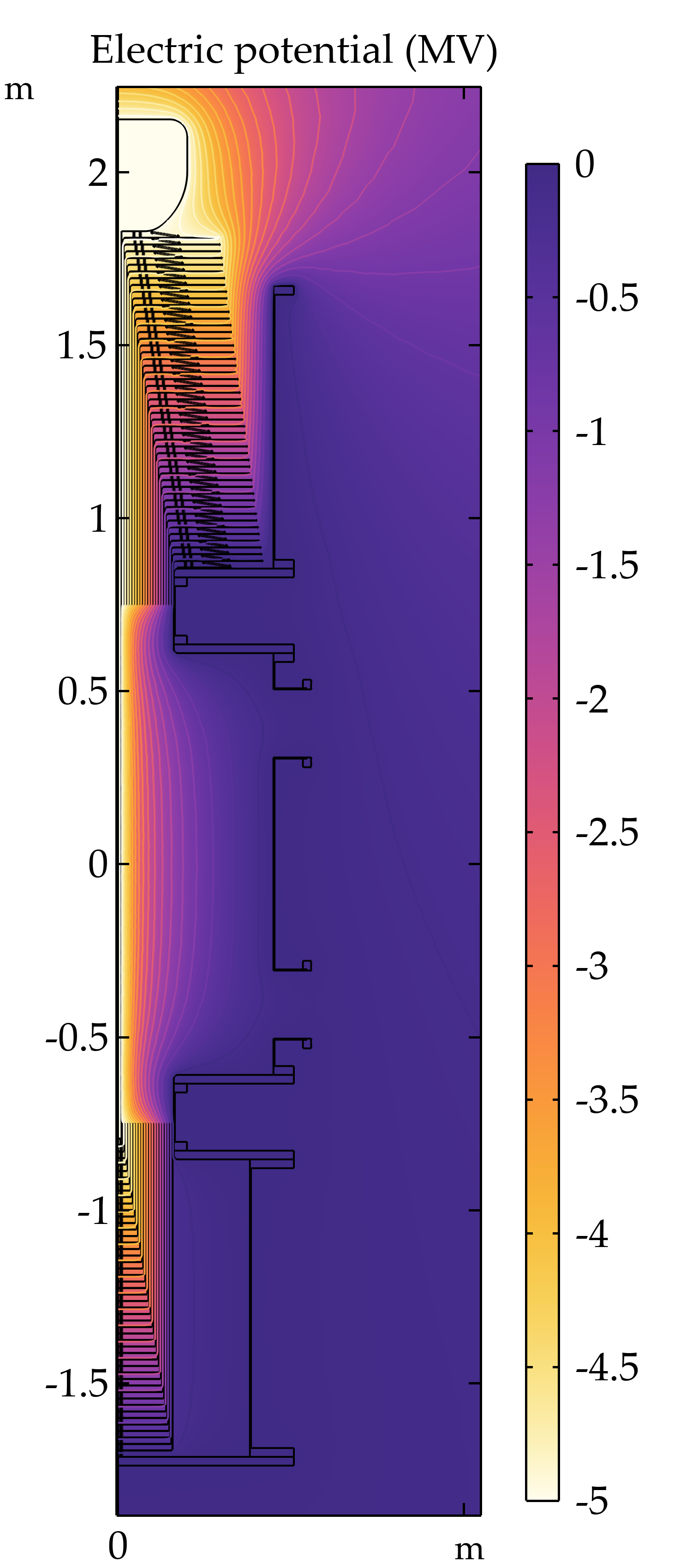}
         \caption{Electric-potential distribution.}
         \label{fig:volt_orbitron}
     \end{subfigure}
     \begin{subfigure}[b]{0.23\textwidth}
         \centering
         \includegraphics[width=\textwidth]{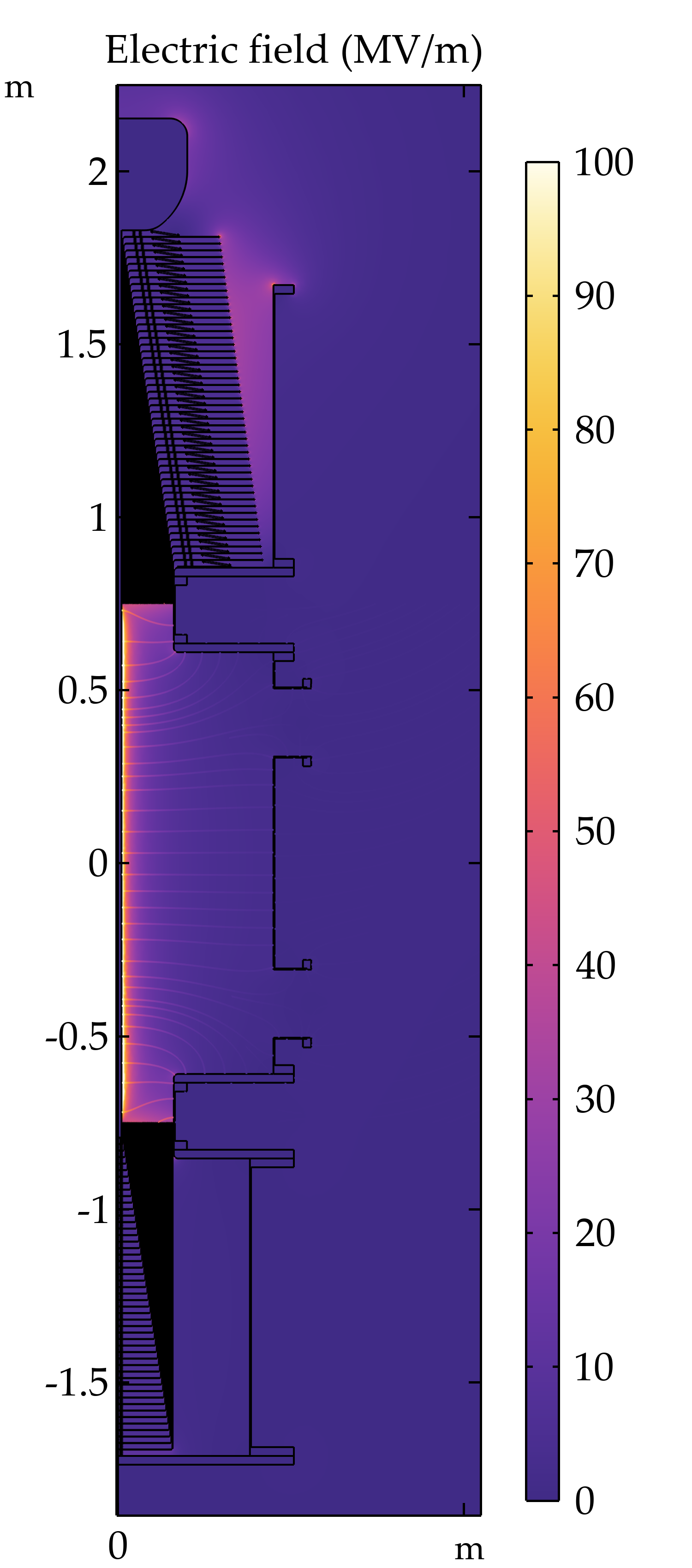}
         \caption{Electric-field distribution.}
         \label{fig:efield_orbitron}
     \end{subfigure}
     \caption{Electrostatic simulation of the bushing and the Orbitron.}
     \label{fig:bushing_n_orbitron}
\end{figure}

The vacuum bushing transmits HV from the pressurized gas vessel to the vacuum vessel, where pressure ranges from $10^{-6}$ to $10^{-1}$~Pa. As shown in Figure \ref{fig:bushing_n_orbitron}, insulating spacers placed between electrodes -- including the central cathode and the chamber -- serve as the interface between the pressurized and vacuum regions while maintaining 100~kV across each spacer. A major concern is flashover along the insulator surface, where the breakdown voltage is typically lower than that of both the bulk solid dielectric and the gaseous or vacuum dielectrics on either side. Although the progression of flashover in vacuum differs from that in pressurized gas, it is commonly agreed that both are initiated by electron emission at the triple junction~\cite{sudarshan_mechanisms_1986,li_surface_2023}, where the cathode, insulator, and vacuum (or gas) meet and experience intense electric stress from the mismatch in relative permittivities and conductivities between the solid and the gaseous or vacuum insulators.

\begin{figure}[h]
    \centering
    \includegraphics[width=0.48\textwidth, angle=0]{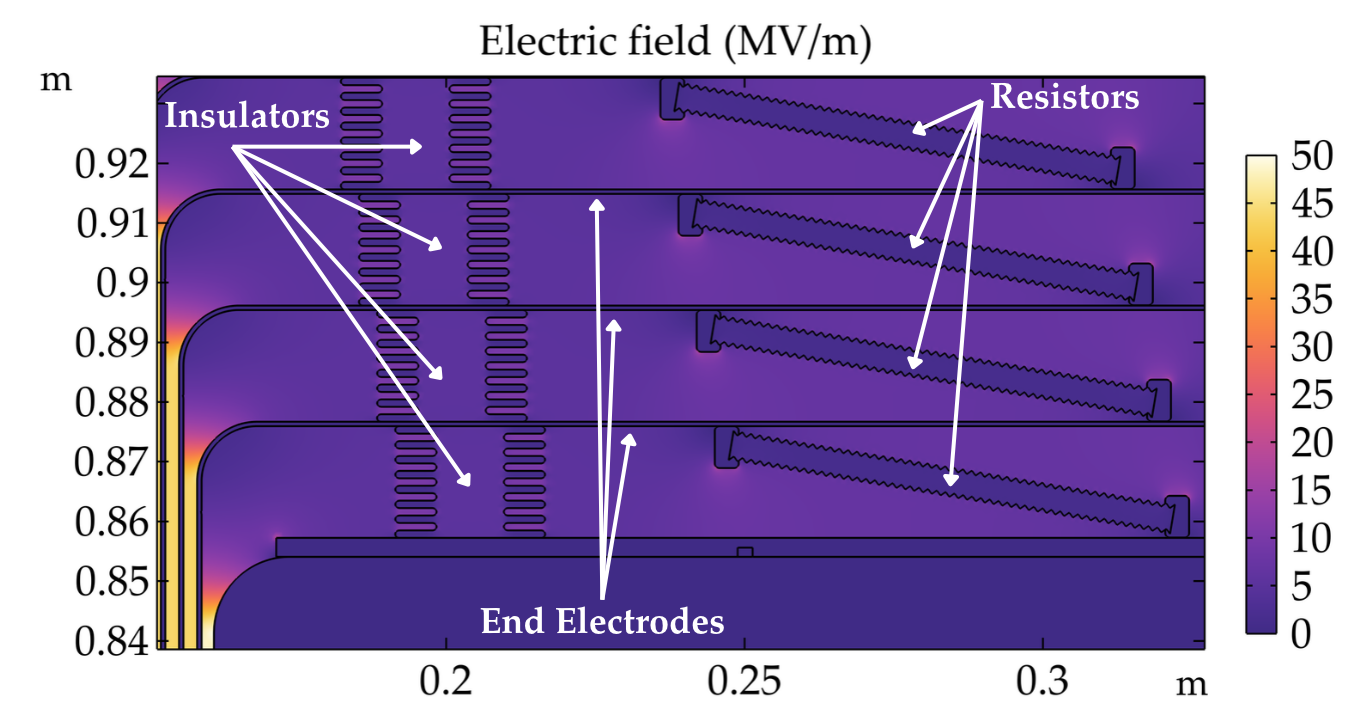}
    \caption{Electric-field distribution near the magnet bore and the first end electrodes, showing the corrugated insulators and resistive voltage-grading elements.}
    \label{fig:insulators}
\end{figure}

This is exactly the failure mode that the Hammerhead prototype was designed around, and its measured triple-junction field of 1.04 MV/m at 300 kV (Section~\ref{sec:lessons}) informs the geometry adopted at each of the 50 gaps here. Figure~\ref {fig:insulators} shows the cathode triple-junction region in the present design. The simulated triple-junction field is $\sim$7.7 MV/m, well below the switch-on field reported for refractory metals such as tungsten or molybdenum~\cite{furuta_reduction_2005}. In addition, the corrugated insulator has a flashover path of 210 mm per 100 kV gap, which is an effective geometry for inhibiting surface flashover in a compact standoff~\cite{borghei_hammerhead_2025,miller_flashover_1993}.

Of the vacuum-compatible ceramics available for the spacers, we selected 97\%-pure aluminum oxide (Al\textsubscript{2}O\textsubscript{3}) because it is widely used and can be brazed directly to tungsten electrodes. The 18~mm-thick insulator provides sufficient solid-dielectric holdoff for the 100~kV gap, given its dielectric strength of $\sim$15.5 MV/m.

Unlike other insulating media, the dielectric strength of a vacuum is strongly dependent on the total gap voltage: breakdown is associated with a progressively lower cathode gradient as the gap distance and voltage increase, a phenomenon known as the ``total voltage'' effect~\cite{trump_insulation_1947}. A single-gap distance exceeding $\sim$3~m would be required to prevent vacuum breakdown at 5 MV~\cite{cranberg_initiation_1952}. To keep the device compact, the 5 MV total potential is divided into fifty 100 kV gaps. For a uniform field, each gap can be as small as 2~mm (corresponding to a breakdown field above 50 MV/m)~\cite{spolaore_large_1997}.

The horizontal segments of the electrodes on the vacuum side (past the insulators) are spaced 18~mm apart, giving an average field of 5.5 MV/m. These L-shaped electrodes then transition to vertical segments inside the magnet bore, where the axial field is strongest, and the electrodes directly face the plasma. In the vertical segments, the inter-electrode clearance is reduced to $\sim$2.5~mm, giving an average field of 40 MV/m. While 40 MV/m is at the high end of what can be tolerated in a purely vacuum gap, the strong axial magnetic field suppresses radial electron motion and provides the magnetic insulation that makes this field level acceptable (see Section~\ref{sec:magnetization} and Ref.~\cite{lovelace_theory_1974}).

Although the intermediate electrodes divide the potential difference across the individual insulating stages, the main Orbitron cell undergoes the full 5 MV potential difference in a single gap between the central cathode and the surrounding anode. In a single gap, the existing clearance is insufficient due to the total voltage effect discussed above. The strong axial magnetic field present in the system suppresses cross-field electron transport and provides magnetic insulation, as detailed later in Section \ref{sec:orbitron_chamber}.

\subsection{Resistive voltage grading}
To maintain a controlled potential distribution along the 5 MV bushing, the 49 intermediate electrodes are connected through a series network of 50 high-voltage resistive elements, as illustrated in Fig. 11. The proposed resistors consist of Al$_2$O$_3$ ceramics doped with ruthenium oxide (RuO$_2$), a commercially developed resistive material whose electrical resistance can be adjusted through the RuO$_2$ concentration. Each resistor incorporates a corrugated surface to increase the creepage distance and is designed for a nominal voltage-withstand rating of 150 kV, providing a margin above the 100 kV operating potential difference between consecutive electrodes.

The resistance is selected to minimize the parasitic current drawn by the grading network while maintaining the intended voltage distribution. For 50 identical resistors connected in series across the full 5 MV potential difference, selecting a resistance of $R=5\,\mathrm{G}\Omega$ per stage yields a total resistance of $250\,\mathrm{G}\Omega$ and a grading current of $I_{\mathrm{div}}=20\,\mu\mathrm{A}$. The resulting power dissipation is about 2 W per resistor, or 100 W for the complete divider, which is 0.1\% of the nominal 100 kW DC output power. \\

\subsection{Orbitron chamber and plasma-load interaction}\label{sec:orbitron_chamber}
The main vacuum chamber houses the Orbitron assembly, which consists of a central rod electrode (cathode), the chamber wall (anode), and end electrodes that provide axial confinement at each longitudinal end. Although these end-electrode assemblies appear identical from a plasma-facing perspective, their structural and electrical integration differ. On the left-hand (HV) side, the electrode stack is brought into vacuum through the brazed-ceramic bushing described above. On the right-hand side, a dividing resistor string along the $z$-axis maintains the electrode standoff; because these electrodes lie entirely within the vacuum vessel, the brazed-ceramic insulators needed for the left-side HV interface are not required.

In the primary plasma region between these electrode sets, the electrical behavior deviates significantly from that of a conventional electrostatic gap. In a cold, unmagnetized vacuum, the system would likely break down before reaching 5 MV. However, a fusion plasma in crossed electric and magnetic fields ($\mathbf{E}\times\mathbf{B}$) fundamentally changes the conduction dynamics. With the axial field ranging from 2.5 T at the device center to 10 T in the magnet bore, electrons are highly magnetized; their mobility along the field lines is orders of magnitude greater than in the transverse direction. This high parallel conductivity keeps the plasma potential nearly constant along each field line, effectively making them equipotentials~\cite{liziakin_generation_2024}. Conversely, electron transport across field lines is severely inhibited, suppressing conventional vacuum-breakdown mechanisms and maintaining the extreme potential gradients. When a bias is applied between electrodes on discrete field lines, the resulting potential difference persists throughout the plasma volume, establishing a stable system of stationary $\mathbf{E}\times\mathbf{B}$ fields. A detailed characterization of these plasma--HV interactions and their transient dynamics will be presented in a subsequent publication.

\subsection{Neutron shielding and radiation protection}

The D-T fusion reactions in the Orbitron generate energetic neutrons that impose additional requirements on the surrounding structures and the high-voltage power supply. For the operating point considered in this study, the MC-TRANS++ model predicts a neutron power of approximately 273 kW and a neutron wall loading of 66.4 kW m\(^{-2}\). Therefore, a neutron blanket and shielding system will be required to manage fusion-neutron energy deposition and limit radiation exposure to surrounding equipment. The detailed design of this system, including its material composition, geometry, and shielding thickness, is beyond the scope of the present high-voltage conceptual design and will require dedicated neutron-transport and thermal analyses.

Radiation protection is particularly important for the PFSCR's semiconductor components, whose performance and reliability may be affected by neutron-induced displacement damage, total ionizing dose, and single-event effects. The proposed integration strategy is to place radiation-sensitive components behind the primary shielding wherever practicable and to incorporate additional local shielding where necessary. The shielding design must therefore account for neutron and secondary gamma-ray transport, including possible streaming paths through structural openings and high-voltage penetrations. Component radiation tolerance and the accumulated radiation dose over the intended operating lifetime will be evaluated as part of the detailed system design.

The gas-insulated high-voltage vessel also requires consideration of radiation-induced degradation of its insulating medium. Neutron and secondary gamma-ray exposure may radiolytically decompose the C$_5$F$_{10}$O/CO$_2$/O$_2$ gas mixture, potentially affecting its dielectric strength, chemical stability, and long-term insulating performance. Consequently, the final shielding configuration must limit the radiation exposure of the gas-insulated power supply to levels compatible with its required operating lifetime. Coupled neutronics calculations and radiation-qualification measurements will be needed to establish the relevant dose levels and verify the dielectric performance of the insulating gas and associated materials under representative radiation conditions.

\section{Conclusion and Future Work}\label{sec:conclusion}
This article presents the conceptual engineering design of a 5 MV high-voltage system for a next-generation, net-positive compact Orbitron fusion device. Building on quantitative lessons from a compact 300 kV prototype (\emph{Hammerhead}), the proposed architecture combines a SiC-MOSFET CLLC resonant tank with a 120-stage parallel-fed series-cascaded rectifier to deliver 5 MV at 20 mA. Key advances compared with the state of the art include: an end-to-end efficiency of $\sim$85\%, versus $\sim$60\% for triode-driven Dynamitrons; an environmentally-sustainable gas insulation strategy using a C\textsubscript{5}F\textsubscript{10}O / CO\textsubscript{2} / O\textsubscript{2} mixture with GWP more than four orders of magnitude lower than SF\textsubscript{6}; a compact vacuum-delivery architecture with triple-junction shielding and corrugated insulator spacers that powers the cathode and distributes voltage uniformly across the end-electrode assemblies; and an integrated $\mathbf{E}\times\mathbf{B}$ insulation strategy that exploits the 2.5--10 T magnetic field to maintain 40 MV/m gap fields inside the plasma region.

According to the 0-D MC-TRANS++ code, a compact Orbitron with a first-wall radius of 45 cm and length of 1 m, operating at a central magnetic field of 2.5 T and an end-coil field of 10 T, could achieve a plasma $Q\approx 8.5$ when driven by the proposed integrated HV system. This corresponds to 433 kW of thermal output, driven primarily by 40 kW delivered to the HV terminal. Depending on the achieved thermal efficiency, such a device could produce a net electrical output of 86--171 kW\textsubscript{e} within a physical form factor of approximately 2.7 m diameter and 6.5 m length.

Future work will focus on (1) subsystem testing and transient analysis to characterize the interaction between the HV supply and the dynamic plasma load, especially during startup and fault modes;
(2) experimental qualification progressing from individual 100 kV bushing stages to intermediate-voltage multi-stage assemblies under representative vacuum and magnetic-field conditions; (3) coupled neutronics and radiation-transport analyses to establish shielding requirements and component radiation exposure; and (4) further efficiency optimization of the HV generation chain.

\section*{Acknowledgments}
The authors thank Steve Battel, Craig Miller, Brian Riordan, and Corentin Lazarus for insightful discussions and their careful review of the manuscript.

\section*{Funding}
This work was funded internally by Avalanche Energy Designs Inc.

\section*{Data Availability}
The data supporting the findings of this study are contained within
the article. Additional data may be made available upon request.

\printcredits

\bibliographystyle{cas-model2-names}
\bibliography{refs}

\end{document}